\documentclass[preprintnumbers,article,amsmath,amssymb,floatfix,10pt,prd,onecolumn,
superscriptaddress,nofootinbib]{revtex4}
\usepackage{bbm}
\usepackage{amsfonts}
\usepackage{mathrsfs}
\usepackage{latexsym}

\usepackage{epsfig}
\usepackage{graphicx}
\usepackage{epstopdf}
\usepackage{placeins}
\usepackage{float}

\usepackage{amssymb}
\usepackage{amsmath}
\usepackage[cp1251]{inputenc}
\usepackage{dcolumn}
\usepackage{bm}
\usepackage{color}
\usepackage{comment}
\usepackage{xcolor}

\begin{document}

\title{\bf Ricci Inverse Anisotropic Stellar Structures}

\author{ M. Farasat Shamir}
\email{farasat.shamir@nu.edu.pk}\affiliation{National University of Computer and
Emerging Sciences,\\ Lahore Campus, Pakistan.}
\author{Mushtaq Ahmad}
\email{mushtaq.sial@nu.edu.pk}\affiliation{National University of Computer and
Emerging Sciences, Islamabad,\\ Chiniot-Faisalabad Campus, Pakistan.}
\author{G. Mustafa}
\email{gmustafa3828@gmail.com: gmustafa3828@zjnu.edu.cn }\affiliation{Department of Physics, Zhejiang Normal University,
Jinhua 321004, China}
\author{Aisha Rashid}
\email{aisharashid987@gmail.com}\affiliation{National University of Computer and
Emerging Sciences,\\ Lahore Campus, Pakistan.}

\begin{abstract}

This paper offers novel quintessence compact relativistic spherically symmetrical anisotropic solutions under the recently developed Ricci inverse gravity \cite{Amendola}, {by employing Krori and Barua gravitational potentials, $Ar^2=\nu(r), ~\&~Br^2+C=\mu(r)$ (with A, B, and C being real constants).} For this objective, a specific explicit equation of state, connecting energy density and radial pressure, i.e., $p_r=\omega\rho$, such that $0<\omega<1$, has been utilized with an anisotripic fluid source. Ricci inverse field equations are used to find the exclusive expressions of the energy density, radial and tangential stresses, and the quintessence energy density, the critical physical attributes reflecting the exceptional conduct of extremely dense matter configuration. For the observatory source stars $Her X-1$, $SAX J 1808.4-3658$ \& $4U 1820-30$, all the important physical quantities like energy densities, tangential and radial pressures, energy conditions, gradients, anisotropy, redshift and mass-radius functions, and stellar compactness have been worked out and analyzed graphically. It has been concluded that all of the stellar formations under consideration remain free from any undesirable central singularity and are stable.
\end{abstract}

\maketitle
\section{Introduction}

The self-gravitation mechanism and its effects have a significant impact on new astrophysics studies  and its background. This gravitational collapse results in the formation of new stellar components  known as compact stars. These compact objects are thought to be exceedingly dense because the  are  massive  but with  smaller volumetric radii. Compact stars have attracted the interest  of researchers in relativistic astronomy due to the relativistic structures, which are well explained by general relativity and by different modified theories \cite{weber,delgo,jimr10,jimr11,jimr12,jimr13,jimr14,jimr15,jimr16,jimr17,ad11,ad8,ad15} as well. Modified and extended theories have been used by countless researchers to overcome the limitations and restrictions of general relativity, which occurred  because of dark matter and dark energies \cite{kra,ost,lid,brat2,wett,ferr} that effects the accelerated universe. Each of these approaches is  without flaws. Some of the researcher have used simple cosmological constant \cite{wein,carrl,brat,cal,armn1,armn,sahn},  which has just recently gained dominance in the present era. Quintessence approach depending upon a scalar field has also been used by \cite{zal,brax,barr11}. The emergence of new dimensional constants is an unappealing feature of most alternative formulations. Another issue is the addition of emerging  fields, either scalar or vector, with no evident relationship to the curvature tensor, which represents a significant change from Einstein's geometrical approach. These difficulties are not necessarily deal breakers, but they are also not grounds in favor of changing Einstein's study of  gravity. Moreover, the Starobinsky models  \cite{staro11,barr33,staro22,barr44,mull11,kosh11,mish11}, which include the $R^2$ corrected term, are one of the most promising models. These models are also the fundamental variants of the $f(R)$ theory  \cite{feli22,nojiri22}. As a result, having a reliable alternative  model may lead to a more revealing picture of our universe's early and late time eras \cite{appl02}. Furthermore, surprising findings may occur in alternate gravitational theories due to the availability of extra $4th$ order corrected  terms like $R^2$ or $R_{\alpha \beta}A R^{\alpha \beta}$.

Deser and Woodard \cite{deser01} investigated two models that allow for a slow reaction to cosmic occurrences while avoiding the normal range. They apply the analysis on FRW space-time by taking the inverse of the D'Alembert operator along with cosmological constant. Soussa and Woodard \cite{sousa} analyzed the altered Ricci scalar, which allowed them to calculate the gravitational effect imparted by the  metric perturbation trace. They have restricted the matter dispersion to have the property that its  gravitational collapse rate  is similar to the rate of space-time extension. Moreover, Amendola et. al. \cite{amen2} have expanded  the idea  of Deser and Woodard by employing the non-local function of altered gravity and changing the Einstein-Hilbert equation with a term $Rf(\square^{-1} R)$, where $f$ is a random function. Furthermore, a non-local altered  gravity \cite{nojiri55,barv} was obtained by adding the expression $m^2 R \Box^{-1} R$, which  has caused a self-accelerating expansion  by establishing the equation of state (EoS) of dark energy, $w_{DE}\simeq -1.14$ \cite{maggi}. Amendola \cite{Amendola} has developed a new theory for gravity known as Ricci inverse (RI) gravity which is dependent on the anti-curvature scalar $A$. In RI model, the term $A$ is the trace of anti-curvature tensor $A^{\alpha \beta}$ that holds the property $A^{\alpha \beta}=R^{-1}_{\alpha \beta}$. Motivated from this recently developed modified theory of gravity, we aim to investigate anisotropic compact structures in RI gravity. For this purpose, we investigate the quintessence structures for compact stars in the RI theory of gravity using $Her X-1$, $SAX J 1808.4-3658$, and $4U 1820-30$ as  star sources. Our work is managed as follows: Section II presents the fundamental concepts, structure, and equations of the RI gravity. In Section III, the matching constraints for the RI gravity  are developed using Schwarzschild's geometry. Section IV is devoted to calculating and analysing physical properties such as energy density, pressure profiles and quintessence density. The final section contains come concluding remarks.

\section{Inverse Ricci Gravity}

The anticurvature tensor, constructed as inverse of Ricci tensor, enables the growth of an alternative gravity theory. A  theory like this will need to be examined in order to determine the prevalence of ghosts-like instabilities and infer the consequences on perturbations like gravitational wave propagation, perturbation growth, or the Newtonian bound. Hence, a theory like this may prove a viable option to be assumed as cosmological model after drawing the general motion equations of the generic function with Lagrangian representing  scalars of the curvature and anticurvature.
It has been demonstrated that in every Lagrangian including a term constituted with some positive or negative exponent of the anticurvature scalar, a generalized no-go theorem prohibits cosmological trajectories from joining a decelerated stage to an accelerated one, ruling out a large group of models. In some especially simple examples of models with no additional dimensional scales, the no-go theorem is demonstrated analytically and mathematically. In this novel theory, the various escape routes from the theorem have been examined, highlighting several (quite complicated) Lagrangians that do so. A thorough examination of such concepts in quest of fresh phenomenology and will happen to be quite fascinating if further investigation is made. A tensor $A^{ab}$ as inverse of $R_{ab}$ is defined as
\begin{equation} \label{14}
A^{ab}R_{bc}=\delta^{a}_{c},
\end{equation}
where $A^{ab}$ the anticurvature tensor. The action for Ricci-inverse gravity is given by \cite{Amendola}
\begin{equation} \label{15}
S_{Action}=\int \sqrt{-g}d^4x(\alpha A+R),
\end{equation}
where $A$ is a trace of $A^{ab}$, which is defined as
\begin{equation} \label{16}
A^{ab}=R^{-1}_{ab}.
\end{equation}
The modified field equations are given as \cite{Amendola}
\begin{equation}\label{1004}
R^{ab}-\frac{1}{2}Rg^{ab}-\alpha A^{ab}-\frac{1}{2}\alpha A g^{ab}+\frac{\alpha}{2}\left(2g^{\varrho a}\nabla_{\alpha}\nabla_{\varrho}A^{\alpha}_{\sigma}A^{b\sigma}-\nabla^2A^{a}_{\sigma}
A^{b\sigma}-g^{ab}\nabla_{\alpha}\nabla_{\varrho}A^{\alpha}_{\sigma}A^{\varrho\sigma}\right)={\mathcal{T}}^{ab}
\end{equation}
where $A^{\alpha}_{\sigma}A^{b\sigma}=A^{\alpha\tau}g_{\tau\sigma}A^{\sigma b}=A^{\alpha\tau}A^{b}_{\tau}=A^{\alpha\sigma}A^{b}_{\sigma}
=A^{b}_{\sigma}A^{\alpha\sigma}$ with $8\pi G=1$. The stress tensor possessing anisotropic matter content is read as
\begin{equation} \label{29}
{\mathcal{T}}_{a b}=(\varrho+p_{t})u_{a}u_{b}-p_{t}g_{a b}+(p_{r}-p_{t})v_{a}v_{b}.
\end{equation}
Here, $u_{a}=e^{\frac{b}{2}}\delta_{a}^{0}$ and $v_{a}=e^{\frac{\lambda}{2}}\delta_{a}^{1}$, ${\mathcal{T}}_{a}^{b}={\mathcal{T}}_{a}^{b}+{\mathcal{D}}_{a}^{b}$, ${\mathcal{D}}_{a}^{b}$ gives the stress tensor for quintessence field equation exhibiting energy density expressed as $\rho_{q}$ and the parameter $w_{q}$ represents quintessence field such that ($-1<w_{q}<-\frac{1}{3}$). Also, $-{\mathcal{D}}_{t}^{t}={\mathcal{D}}_{r}^{r}=-\rho_{q}$, and $\frac{(3 w_{q}+1)\rho_{q}}{2}={\mathcal{D}}_{\theta}^{\theta}={\mathcal{D}}_{\phi}^{\phi}$. The spherically symmetric metric is given by
\begin{equation} \label{30}
ds^{2}=e^{\mu(r)}dt^{2}-e^{\nu(r)} dr^{2}-r^{2} d\theta^{2}-r^{2}\sin^{2}\theta d\phi^{2},
\end{equation}
{Now we can choose the Krori and Barua gravitational potentials, which are defined as
\begin{equation} \label{31}
Ar^2=\nu(r)\;\;\;\;\;\;\;\;\;\;\;\;\;\;Br^2+C=\mu(r),
\end{equation}
with $A$, $B$, and $C$ are some arbitrary constants.} By using Eqs. (\ref{29}-\ref{31}) in Eq. (\ref{1004}) with quintessence field, we get the following system of modified field equations
\begin{eqnarray}
\rho +\rho_q &=&e^{-A r^2} \left(\frac{\alpha  \left(\chi _2+\chi _3-\chi _4\right) e^{2 A r^2}}{B^2 \left(r^2 (B-A)+3\right)^4}+\frac{1}{2} \alpha  \chi _1 e^{2 A r^2}-\frac{2 A r^2+e^{A r^2}-1}{r^2}\right), \label{3022}\\
p_r-\rho_q&=& \frac{1}{2} \alpha  \left(\frac{2 \chi _6}{r \left(B (B r+1)-A \left(B r^2+2\right)\right)^4}+\chi _5\right),\label{3023}\\
p_t+\frac{1}{2}  (3w_q+1)\rho_q&=&-r^2 \left(-\alpha  e^{A r^2} \left(\frac{\chi _9}{\left(r^2 (A-B)+e^{A r^2}-1\right)^4}+\chi _8\right)+\alpha  \chi _7 e^{A r^2}+\chi _{10} e^{-A r^2}\right)\label{3024}.
\end{eqnarray}
where $\chi _i$, $\{i=1,...,9\}$ are given in the Appendix (\textbf{I}).
{To find the explicit expressions for the above field equations, we assume a relation between radial pressure and
energy density via $EoS$ parameter, which is defined as
\begin{equation}\label{35}
p_r=\omega  \rho, \;\;\;\;\;\;\;\;\;\;\;0<\omega<1.
\end{equation}
The above relation by Eq. (\ref{35}) is the particular form of $EoS$, the general form of this equation was presented by Herrera and
Barreto \cite{jimr1}. This linear of equation state, though a simplest choice, has been extensively considered to discuss the physical properties of astrophysical compact objects \cite{jimr2,jimr3,jimr4,jimr5,jimr6,jimr7,jimr8,jimr9}.} By using Eq. (\ref{35}) with Eqs. (\ref{3022}-\ref{3024}), we get the final expressions for energy density $\rho$, quintessence energy density $\rho_{q}$, and pressure components $p_{r},\;p_{t}$ as
\begin{eqnarray}
\rho &=&\frac{-\frac{1}{2} \alpha  \left(2 \Psi _1 \Psi _2\right)-e^{-A r^2} \left(\frac{\alpha  \Psi _4 e^{2 A r^2}}{B^2 \left(r^2 (B-A)+3\right)^4}+\frac{1}{2} \alpha  \Psi _3 e^{2 A r^2}-\frac{2 A r^2+e^{A r^2}-1}{r^2}\right)}{\omega +1}, \label{32}\\
\rho_q &=&\frac{e^{-A r^2} \left(-\frac{\alpha  \Psi _4 e^{2 A r^2}}{B^2 \left(r^2 (B-A)+3\right)^4}-\frac{1}{2} \alpha  \Psi _3 e^{2 A r^2}+\frac{2 A r^2+e^{A r^2}-1}{r^2}\right)-\frac{1}{2} \alpha  \left(2 \Psi _1 \Psi _2\right)}{\omega +1}\nonumber\\&+&e^{-A r^2} \left(\frac{\alpha  \Psi _4 e^{2 A r^2}}{B^2 \left(r^2 (B-A)+3\right)^4}+\frac{1}{2} \alpha  \Psi _3 e^{2 A r^2}-\frac{2 A r^2+e^{A r^2}-1}{r^2}\right), \label{32}\\
p_r &=& \frac{\omega  \left(-\frac{1}{2} \alpha  \left(2 \Psi _1 \Psi _2\right)-e^{-A r^2} \left(\frac{\alpha  \Psi _4 e^{2 A r^2}}{B^2 \left(r^2 (B-A)+3\right)^4}+\frac{1}{2} \alpha  \Psi _3 e^{2 A r^2}-\frac{2 A r^2+e^{A r^2}-1}{r^2}\right)\right)}{\omega +1},\label{33}\\
p_t &=&-r^2 \left(e^{-A r^2} \left(\frac{\alpha  e^{2 A r^2}}{r^2 (A-B)+e^{A r^2}-1}+\frac{A-2 B}{r^2}+B (A-B)\right)-\alpha  e^{A r^2}\right.\nonumber\\&\times&\left. \left(\Psi _6 \left(\Psi _8 e^{A r^2}+\Psi _7\right)\right)+\alpha  \Psi _5 e^{A r^2}\right)-\frac{1}{2} \rho_{q} (3 w_{q}+1),\label{34}.
\end{eqnarray}
where where $\Psi _i$, $\{i=1,...,10\}$ are given in the Appendix (\textbf{I}).

\section{Matching Conditions}

The interior border metric does not vary irrespective of the star's geometrical structure, whether from the outside or the inside. Regardless of the reference frame, this emergent circumstance mandates that the metric components must be continuous to the boundary.
While evaluating Schwarzschild's solution connected with stellar remnants in general relativity, it is well-thought-out to be the main priority out of all the accessible different alternatives of matching circumstances. Furthermore, while engaging with modified gravity theories, it is a good idea to account for quasi pressure and energy density.
Many scholars have done outstanding work on boundary conditions \cite{A,B}. By integrating some unique constraints to stellar compact structures as well as the thermodynamically related characteristics, Goswami et al. \cite{G} figured out the matching bounds when exploring the modified gravity.
However, the well-known Schwarzschild's geometry can be used to solve the various stellar solutions. A few constraints are imposed at the boundary $r=R$, that is, $p_r(r=R)=0$, to get the expressions for the field equations.
To get the quantities of the associated unknowns, we must first figure out how to align the internal geometry with the external spacetime. We may employ the Schwarzschild spacetime for this, which is described as
\begin{equation}\label{18}
ds^2=\left(1-\frac{2M}{r}\right)dt^2-\left(1-\frac{2M}{r}\right)^{-1}dr^2-r^2\left(d\theta^2+\sin^2\theta d\phi^2\right).
\end{equation}
The total mass trapped within the star is denoted by $M$, while the radius of the compact celestial leftover is denoted by $R$. {It is important to mention here that whatever the geometry of the star is, either derived internally or
externally, the boundary metric must remains the same with the condition that the components of the metric tensor irrespective of
the coordinate system  across the surface of the boundary will
remain continuous. No doubt, in theory of general relativity, the Schwarzschild solution has been really important in motivating us to choose from the diverse possibilities of the matching conditions. However,  while investigating the compact objects in some alternative gravity theories, modified Tolman-Oppenheimer-Volkoff equations with zero pressure and energy density, the solution outside the star can differ from Schwarzschild's solution. Thus,
it is expected that the solutions of the modified Tolman-Oppenheimer-Volkoff equations with non-zero energy density and pressure may accommodate Schwarzschild's geometry with some specific choice of modified gravity model \cite{R1, R2, R3, R4,R5}. Perhaps this might be the reason or justification that Birkhoff's theorem may not hold in alternative theories of gravity. The detailed investigation on the topic in the context of RI gravity can be an interesting future task.}\\\\
By examining the metric potentials, the accompanying expressions are obtained at $r=R$
\begin{eqnarray}
&&g_{tt}^+= g_{tt}^-,\;\;\;\;\;\;\;g_{rr}^+= g_{rr}^-,~~~~\frac{\partial g_{tt}^+}{\partial r}=\frac{\partial g_{tt}^- }{\partial r}.\label{26}
\end{eqnarray}
The inner and outer signatures respective to the boundary $r=R$ are reflected as (-) and (+), respectively. Comparison of the inner and outer matrices, the values of the desired unknown quantities can be determined as follows:
\begin{eqnarray}
&&A=-\frac{log(1-\frac{2M}{R_{\varepsilon}})}{R_{\varepsilon}^2},~~~~B=-\frac{Mlog(1-\frac{2M}{R_{\varepsilon}})^{-1}}{R_{\varepsilon}^2}\label{28},
~~~C=log(1-\frac{2M}{R_{\varepsilon}})-\frac{M}{R_{\varepsilon}(1-\frac{2M}{R_{\varepsilon}})}.
\end{eqnarray}
The quantities $M$, $R$, $M/R$, and the calculated values of the involved parameters like $A$ and $B$, are provided in Tab. \ref{tab1}. Actually, the values of the parameters $A$ and $B$ come from the Eq. \ref{28}. As it's very clear from the Eq. \ref{28}, there is no effect of the quintessence field $w_q$. In fact, Eq. \ref{28} only depends upon the observational values of stars like $Her X-1$, $SAX J 1808.4-3658$ \& $4U 1820-30$. Further, it is necessary to mention that quintessence field $w_q$ has an important role in the anisotropic matter, especially in the tangential pressure component.
\begin{table}[h]
\caption{\label{tab1}{Approximated values of involved parameters with $(\alpha =1.3\times 10^{-11})$ and $w_{q}=-0.4;)$  .}}
\vspace{0.6cm}\begin{tabular}{|c|c|c|c|c|c|}
\hline
$Strange Quark Star$ &\textbf{$M$} &\textbf{$R(km)$} &\textbf{$\frac{M}{R}$} &\textbf{$A(km^{-2})$} &\textbf{$B(km^{-2})$} \\
\hline
$Her X-1$&$0.88M_\odot$&$7.7$&$0.168$&$0.006906276428$&$0.004267364618$ \\
$SAX J 1808.4-3658$&$1.435M_\odot$&$7.07$&$0.299$&$0.01823156974$&$0.01488011569$ \\
$4U 1820-30$&$2.25M_\odot$&$10.0$&$0.332$&$0.01090644119$&$0.009880952381$\\
\hline
\end{tabular}
\end{table}
\begin{figure}
\centering \epsfig{file=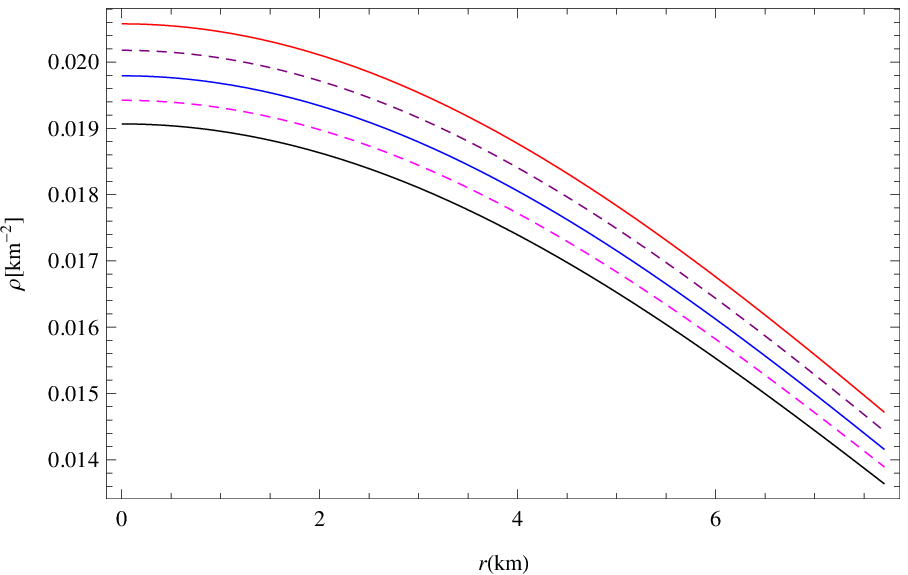, width=.32\linewidth,
height=1.50in} \epsfig{file=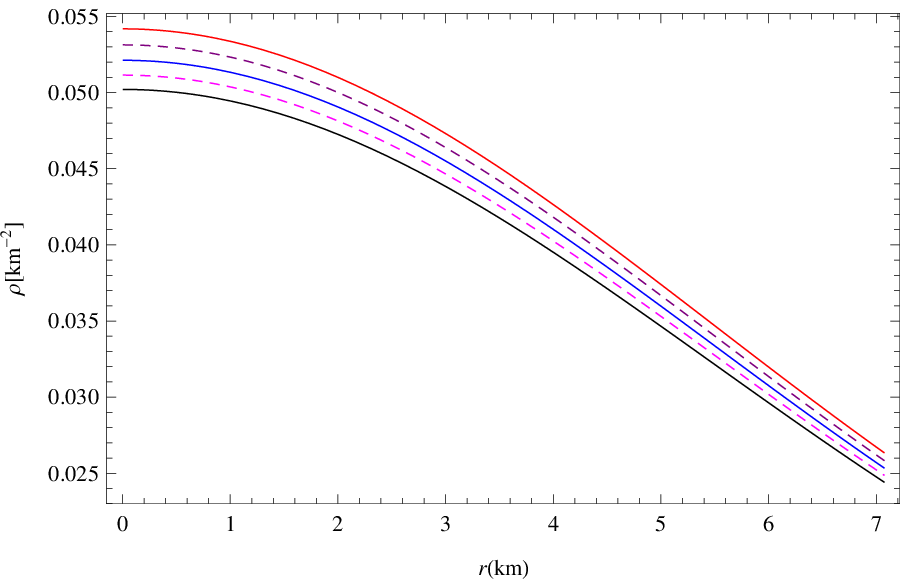, width=.32\linewidth,
height=1.50in}\epsfig{file=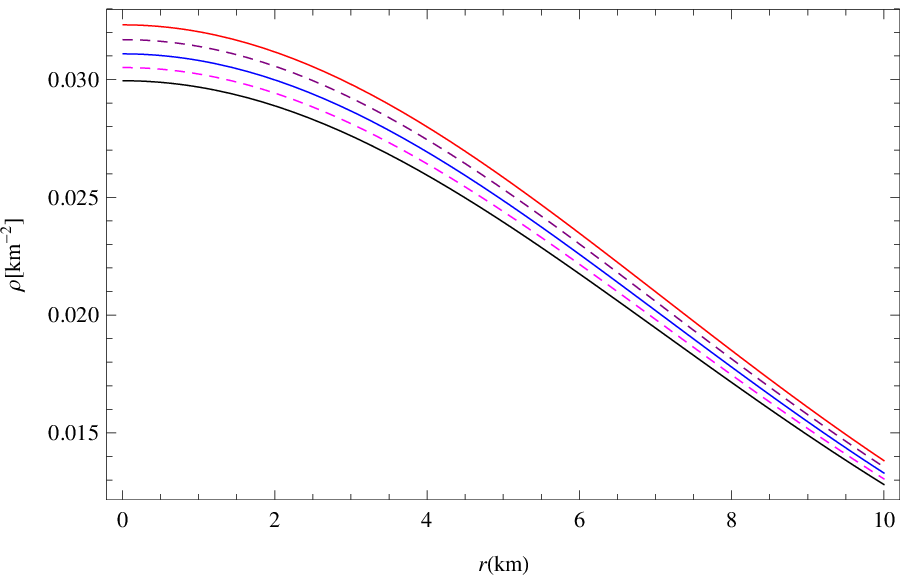, width=.32\linewidth,
height=1.50in}\caption{\label{Fig.1} Shows the energy densities for $Her X-1(Left)$, $SAX J 1808.4-3658(Middle)$, and $4U 1820-30(Right)$ with $\omega=0.01(\textcolor{red}{\bigstar})$, $\omega=0.03(\textcolor{purple}{\bigstar})$, $\omega=0.05(\textcolor{blue}{\bigstar})$, $\omega=0.07(\textcolor{magenta}{\bigstar})$, and $\omega=0.09(\textcolor{black}{\bigstar})$.}
\end{figure}

\begin{figure}
\centering \epsfig{file=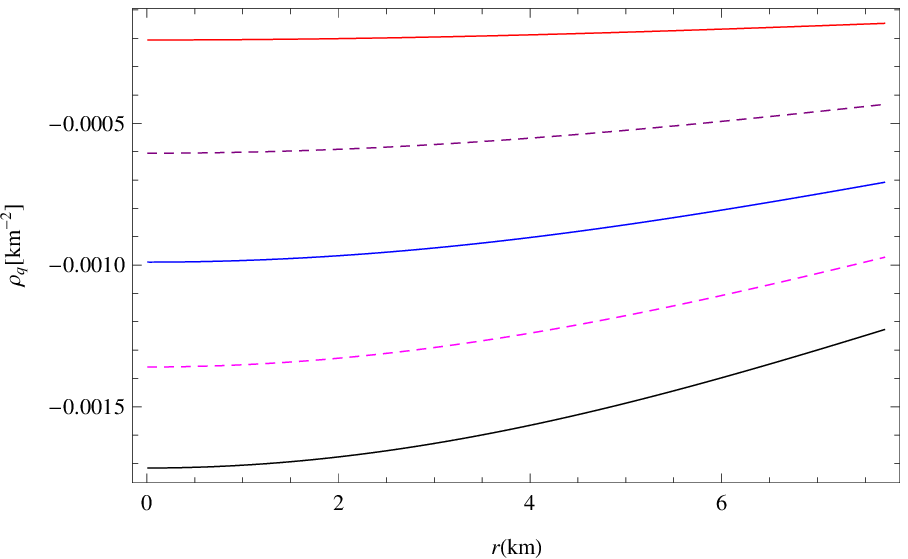, width=.32\linewidth,
height=1.50in} \epsfig{file=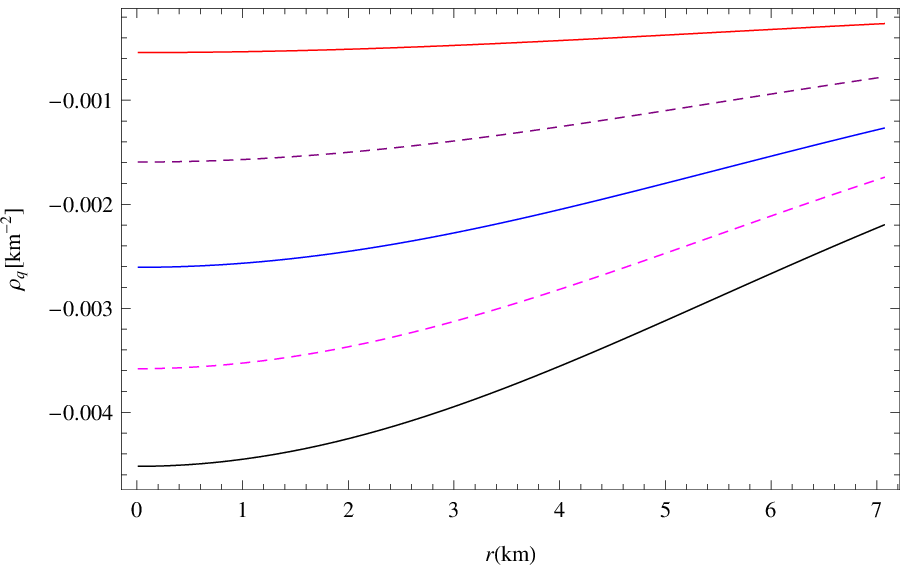, width=.32\linewidth,
height=1.50in}\epsfig{file=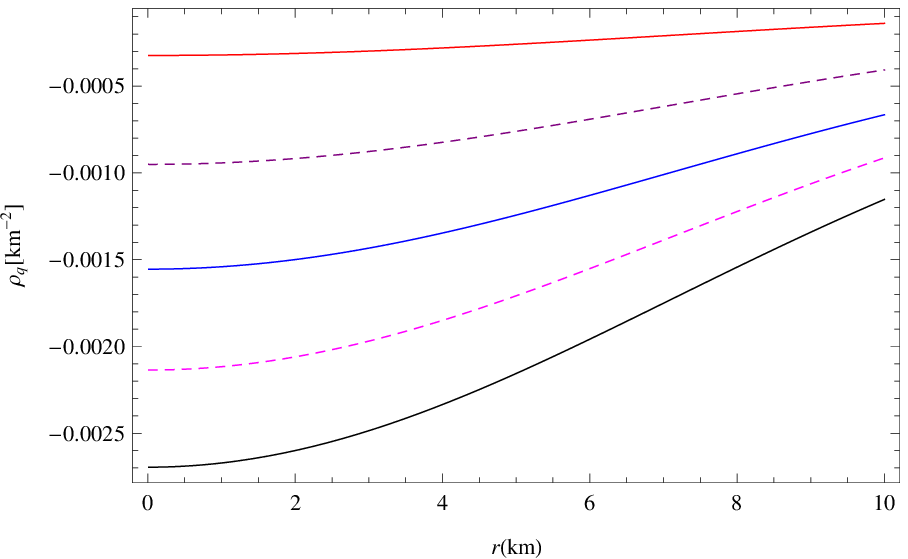, width=.32\linewidth,
height=1.50in}\caption{\label{Fig.2} Shows the quintessence energy density for $Her X-1(Left)$, $SAX J 1808.4-3658(Middle)$, and $4U 1820-30(Right)$ with $\omega=0.01(\textcolor{red}{\bigstar})$, $\omega=0.03(\textcolor{purple}{\bigstar})$, $\omega=0.05(\textcolor{blue}{\bigstar})$,
$\omega=0.07(\textcolor{magenta}{\bigstar})$, and $\omega=0.09(\textcolor{black}{\bigstar})$.}
\end{figure}

\begin{figure}
\centering \epsfig{file=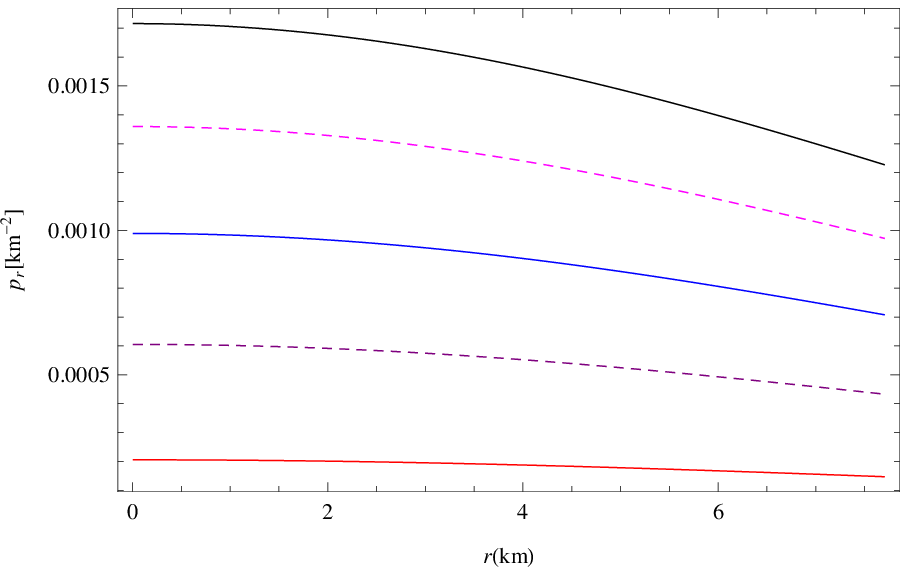, width=.32\linewidth,
height=1.50in} \epsfig{file=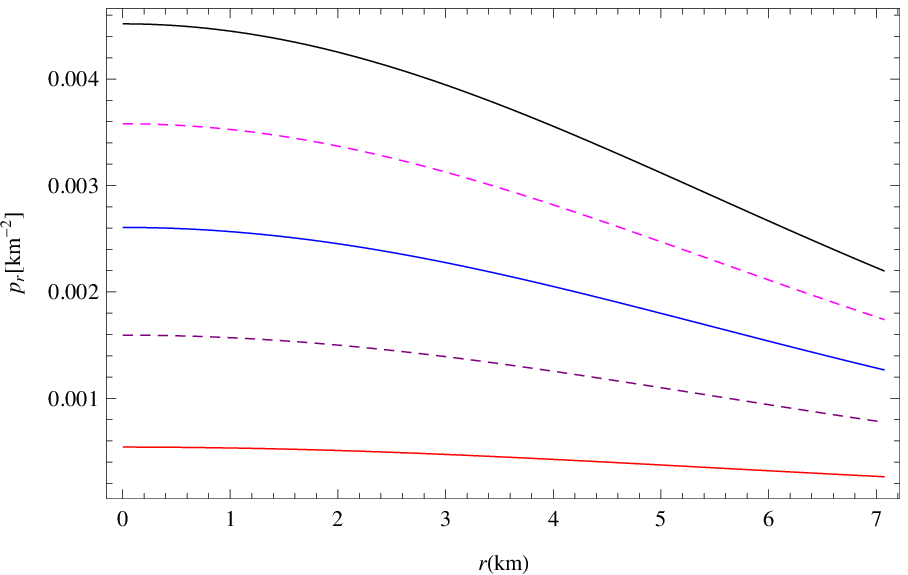, width=.32\linewidth,
height=1.50in}\epsfig{file=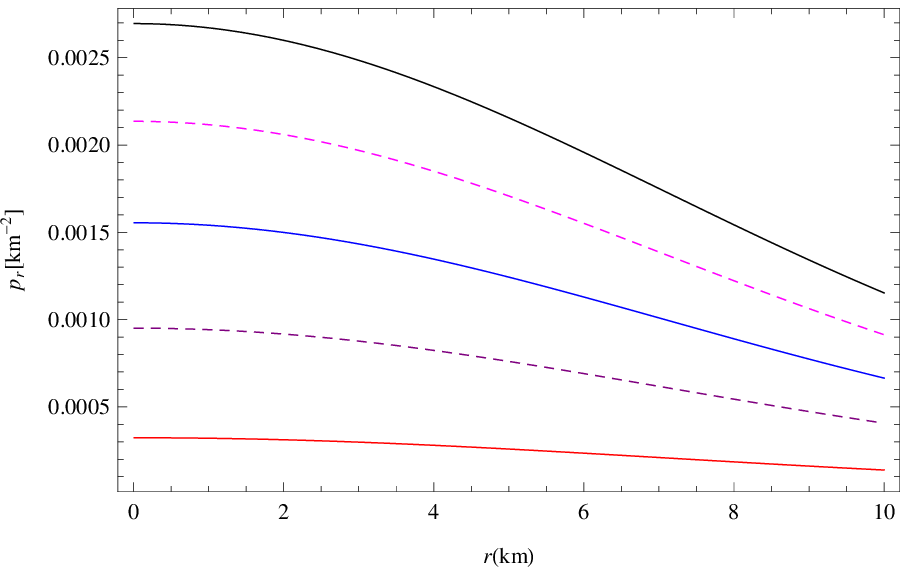, width=.32\linewidth,
height=1.50in}\caption{\label{Fig.3} Shows the radial pressure for $Her X-1(Left)$, $SAX J 1808.4-3658(Middle)$, and $4U 1820-30(Right)$ with $\omega=0.01(\textcolor{red}{\bigstar})$, $\omega=0.03(\textcolor{purple}{\bigstar})$, $\omega=0.05(\textcolor{blue}{\bigstar})$,
$\omega=0.07(\textcolor{magenta}{\bigstar})$, and $\omega=0.09(\textcolor{black}{\bigstar})$.}
\end{figure}

\begin{figure}
\centering \epsfig{file=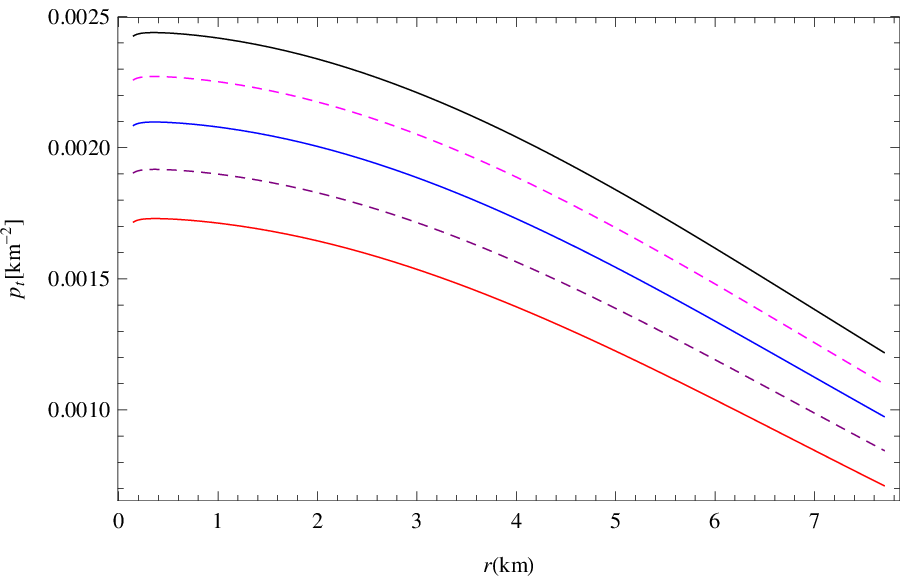, width=.32\linewidth,
height=1.50in} \epsfig{file=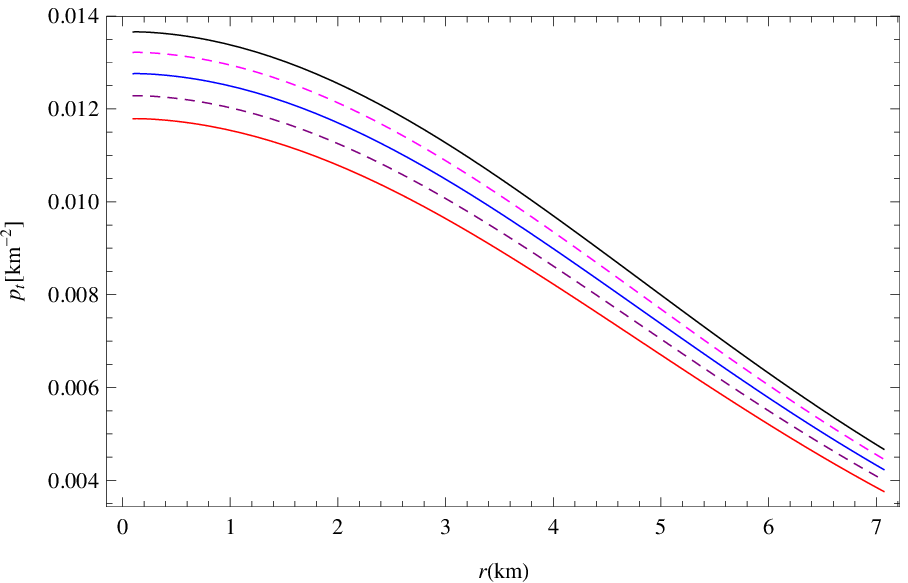, width=.32\linewidth,
height=1.50in}\epsfig{file=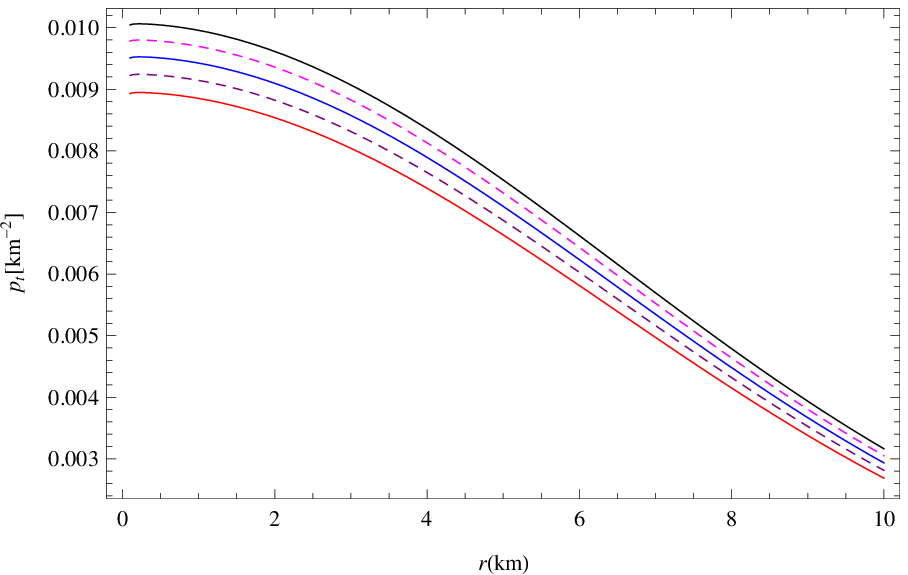, width=.32\linewidth,
height=1.50in}\caption{\label{Fig.4} Shows the tangential pressure for $Her X-1(Left)$, $SAX J 1808.4-3658(Middle)$, and $4U 1820-30(Right)$ with $\omega=0.01(\textcolor{red}{\bigstar})$, $\omega=0.03(\textcolor{purple}{\bigstar})$, $\omega=0.05(\textcolor{blue}{\bigstar})$,
$\omega=0.07(\textcolor{magenta}{\bigstar})$, and $\omega=0.09(\textcolor{black}{\bigstar})$.}
\end{figure}

\section{Physical Analysis}
We devote this section to investigate a few key characteristics associated with considered compact stars. Energy density $\rho$, pressure profiles $p_r$, and $p_t$, and debates on the quintessence field, as well as their physical interpretation under RI gravity, are among them. The energy bounds, anisotropic stress, compactness factor, and the star's sound speed concerning the radial and tangential components are also discussed.
\subsection{Density and Pressure Profiles}
The related evolutions of the energy density, as well as the radial and tangential pressures, make up the most significant stellar environment required for the existence of compact structures. The energy density profiles, and the evolutions of quintessence density, and pressure terms have been studied for $Her X-1$ (Left), $SAX J 1808.4-3658$ (Middle), and $4U 1820-30$ (Right) for diversified values of $\omega=0.01$, $0.03$, $0.05$, $0.07$, $0.09$. The plots of the related graphs as shown in Figs. (\ref{Fig.1}-\ref{Fig.4}) clearly demonstrate the related physical analysis.  The energy density reaches its maximum value near the star's core, indicating the star's super-dense nature. Furthermore, the tangential and radial pressure profiles exhibit positive behavior and reach their maximal values at the surface of star. For our model under RI gravity, the evolutions of the stars under study also imply the existence of an anisotropic structure independent of any possible singularities.

\subsection{Energy Bounds}
\begin{figure}
\centering \epsfig{file=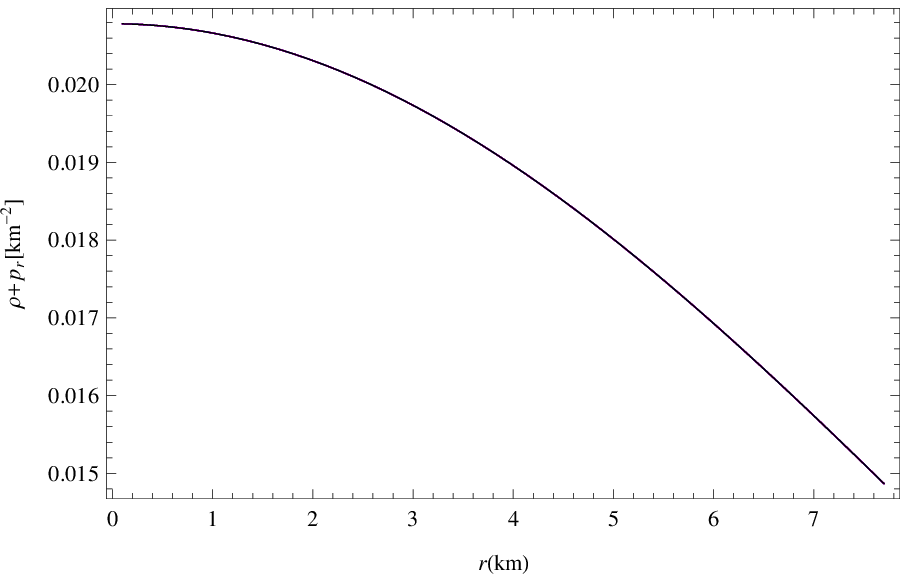, width=.32\linewidth,
height=1.50in} \epsfig{file=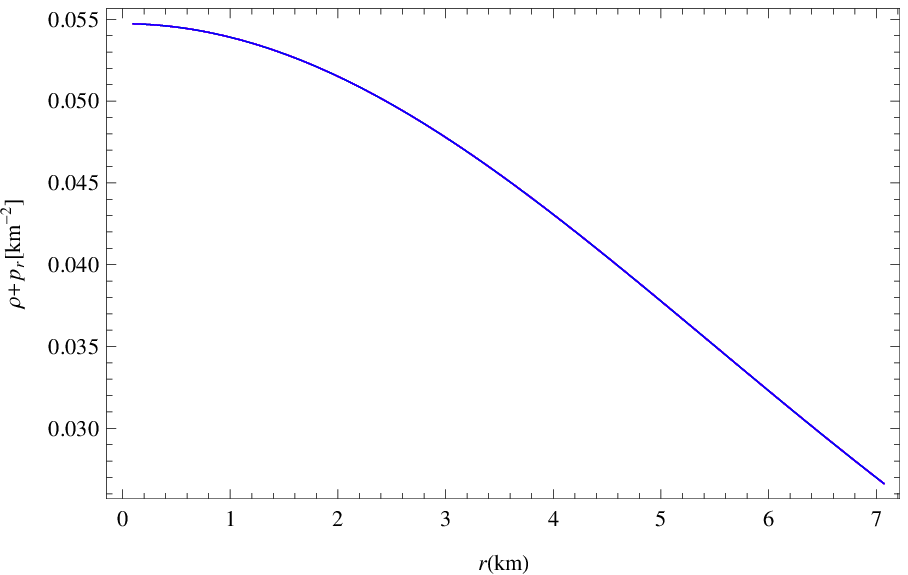, width=.32\linewidth,
height=1.50in}\epsfig{file=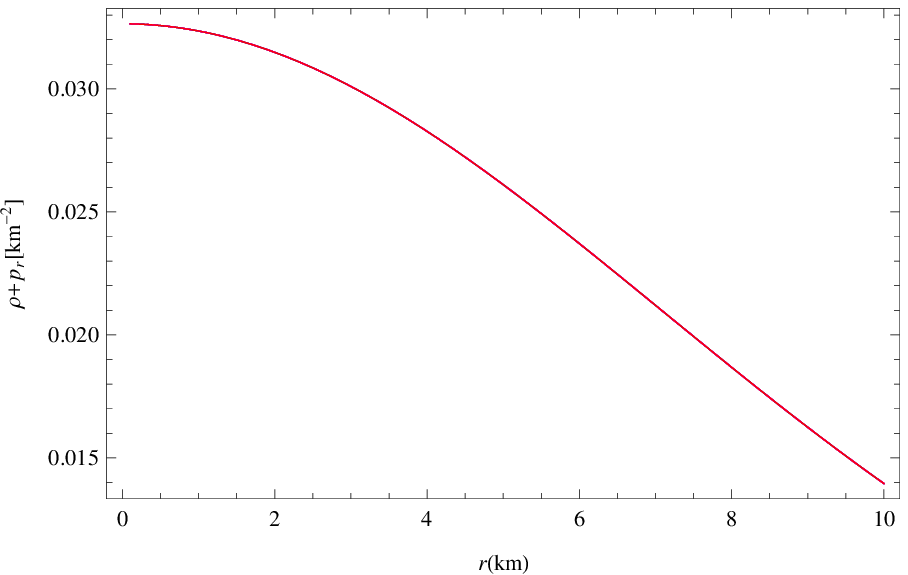, width=.32\linewidth,
height=1.50in}\caption{\label{Fig.9} Shows the energy condition for $Her X-1(Left)$, $SAX J 1808.4-3658(Middle)$, and $4U 1820-30(Right)$ with $\omega=0.01(\textcolor{red}{\bigstar})$, $\omega=0.03(\textcolor{purple}{\bigstar})$, $\omega=0.05(\textcolor{blue}{\bigstar})$, $\omega=0.07(\textcolor{magenta}{\bigstar})$, and $\omega=0.09(\textcolor{black}{\bigstar})$.}
\end{figure}
\begin{figure}
\centering \epsfig{file=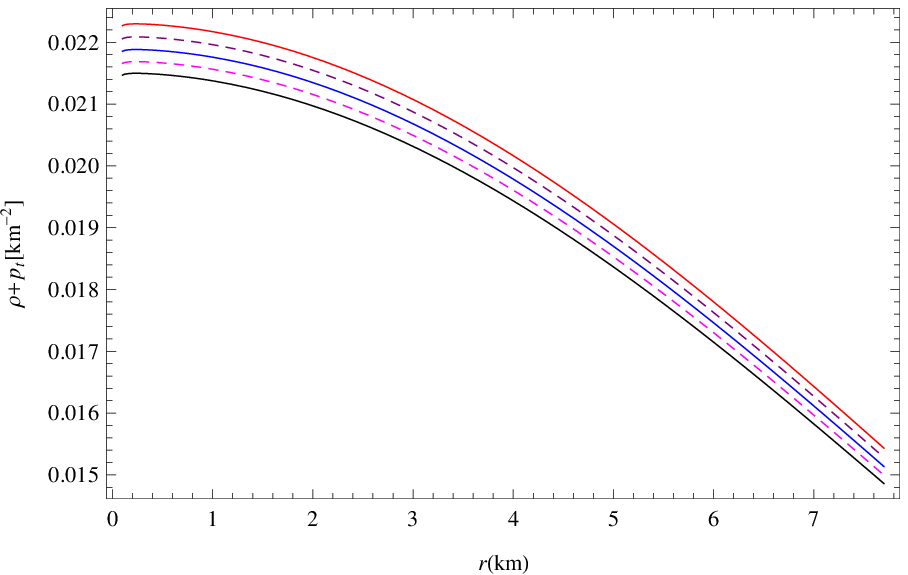, width=.32\linewidth,
height=1.50in} \epsfig{file=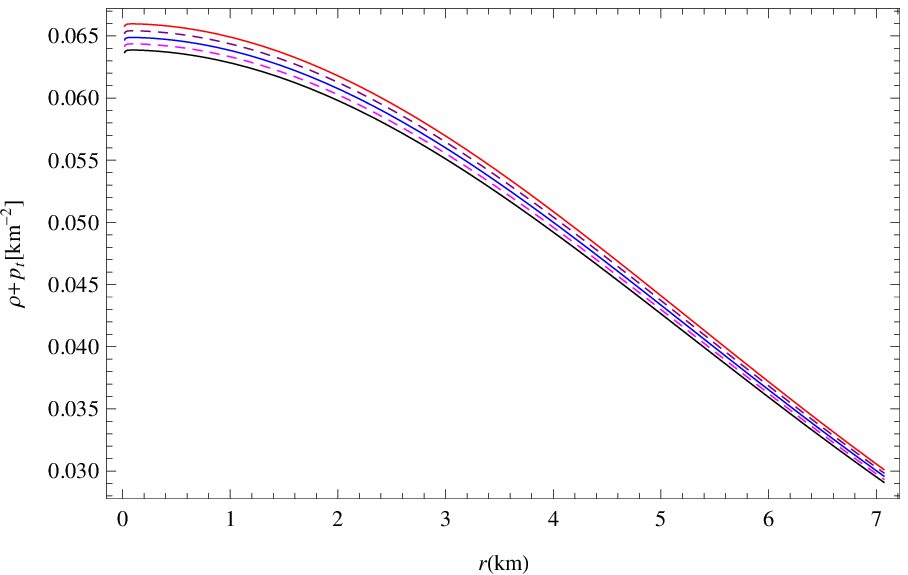, width=.32\linewidth,
height=1.50in}\epsfig{file=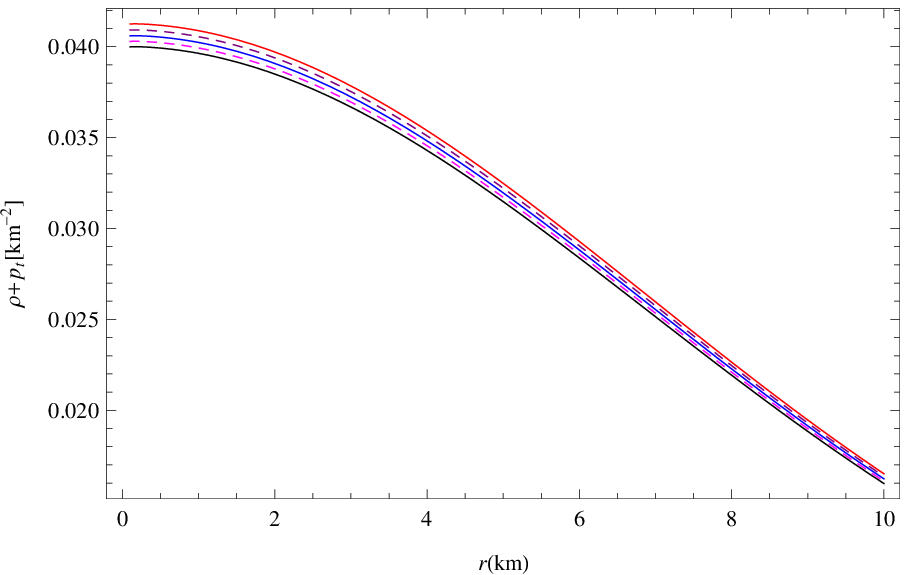, width=.32\linewidth,
height=1.50in}\caption{\label{Fig.10} Shows the energy condition for $Her X-1(Left)$, $SAX J 1808.4-3658(Middle)$, and $4U 1820-30(Right)$ with $\omega=0.01(\textcolor{red}{\bigstar})$, $\omega=0.03(\textcolor{purple}{\bigstar})$, $\omega=0.05(\textcolor{blue}{\bigstar})$, $\omega=0.07(\textcolor{magenta}{\bigstar})$, and $\omega=0.09(\textcolor{black}{\bigstar})$.}
\end{figure}

\begin{figure}
\centering \epsfig{file=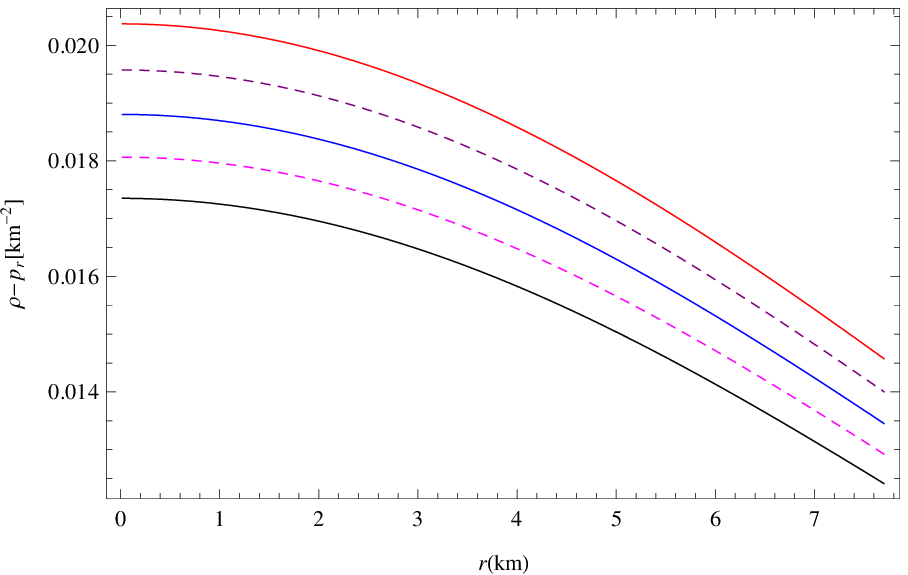, width=.32\linewidth,
height=1.50in} \epsfig{file=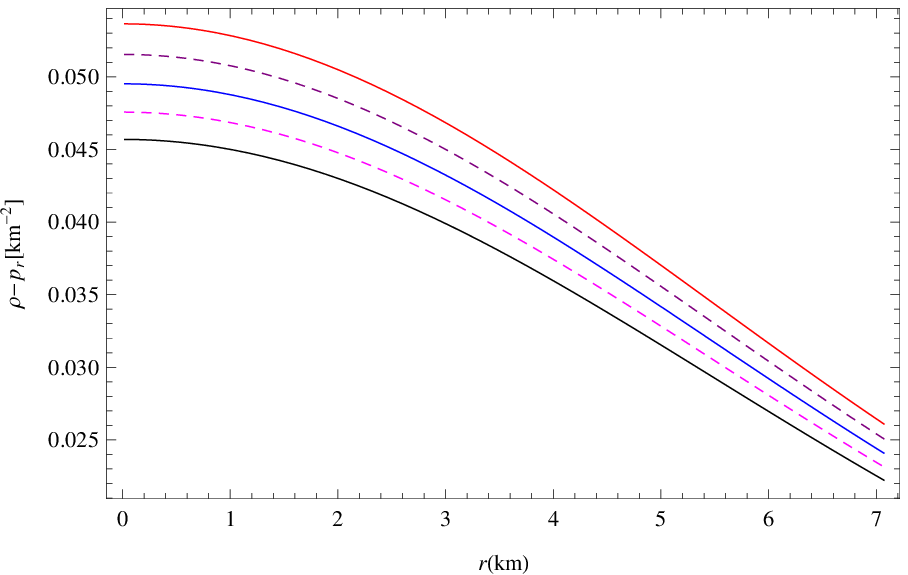, width=.32\linewidth,
height=1.50in}\epsfig{file=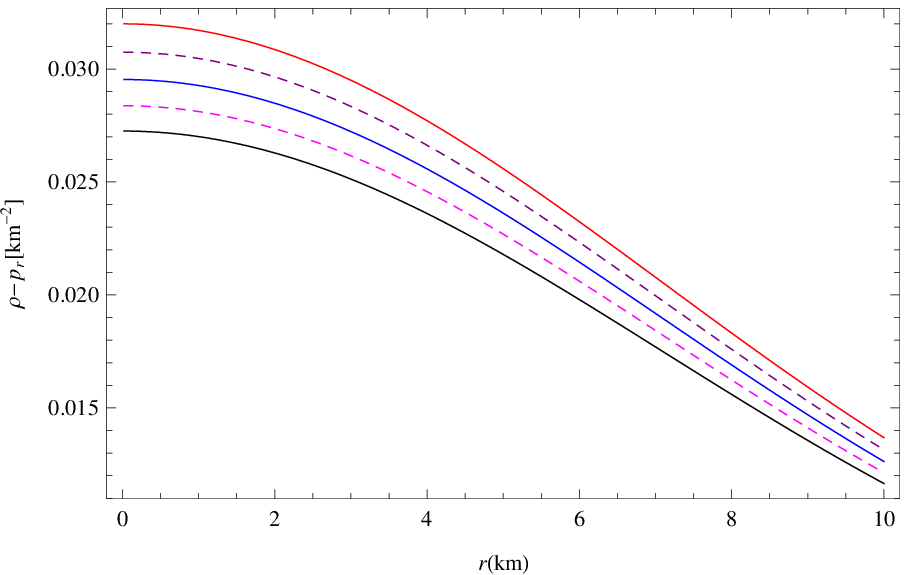, width=.32\linewidth,
height=1.50in}\caption{\label{Fig.11} Shows the energy condition for $Her X-1(Left)$, $SAX J 1808.4-3658(Middle)$, and $4U 1820-30(Right)$ with $\omega=0.01(\textcolor{red}{\bigstar})$, $\omega=0.03(\textcolor{purple}{\bigstar})$, $\omega=0.05(\textcolor{blue}{\bigstar})$, $\omega=0.07(\textcolor{magenta}{\bigstar})$, and $\omega=0.09(\textcolor{black}{\bigstar})$.}
\end{figure}

\begin{figure}
\centering \epsfig{file=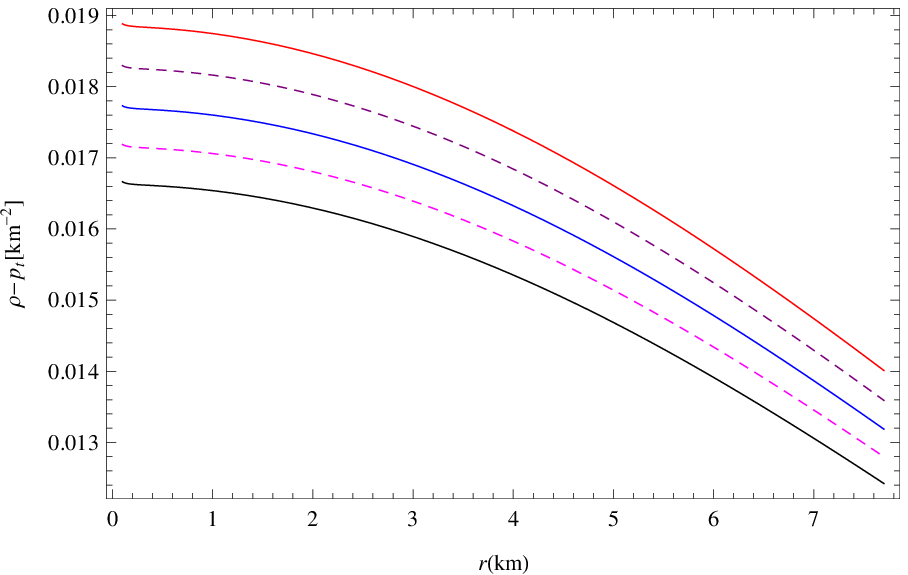, width=.32\linewidth,
height=1.50in} \epsfig{file=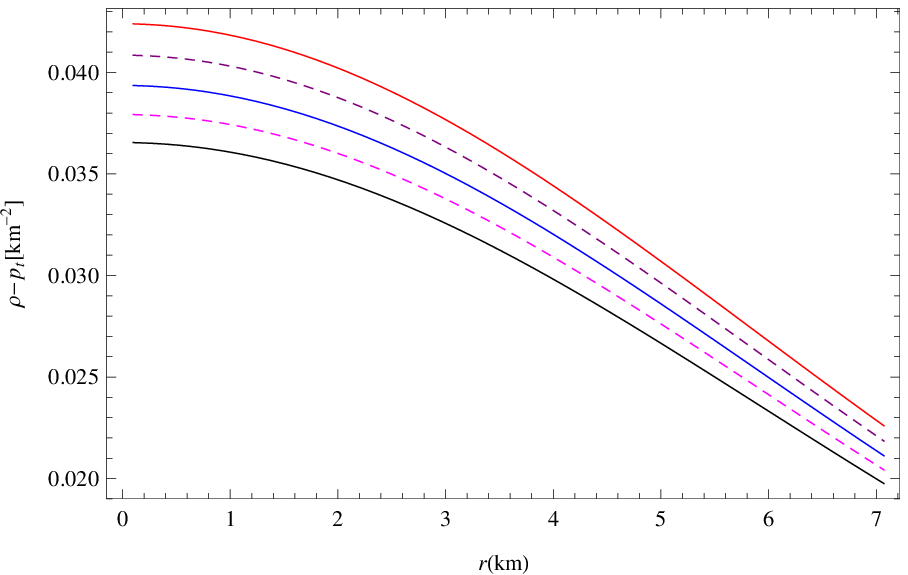, width=.32\linewidth,
height=1.50in}\epsfig{file=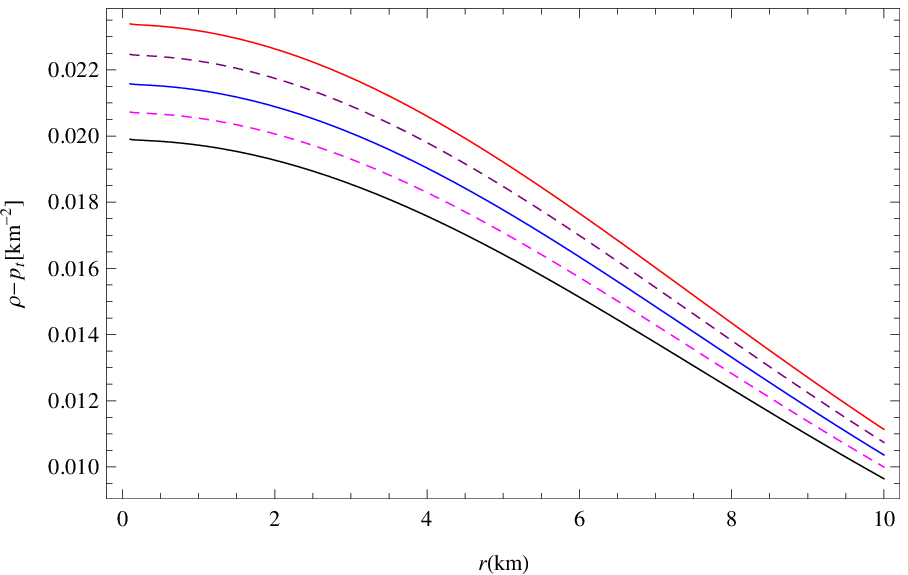, width=.32\linewidth,
height=1.50in}\caption{\label{Fig.12} Shows the energy condition for $Her X-1(Left)$, $SAX J 1808.4-3658(Middle)$, and $4U 1820-30(Right)$ with $\omega=0.01(\textcolor{red}{\bigstar})$, $\omega=0.03(\textcolor{purple}{\bigstar})$, $\omega=0.05(\textcolor{blue}{\bigstar})$, $\omega=0.07(\textcolor{magenta}{\bigstar})$, and $\omega=0.09(\textcolor{black}{\bigstar})$.}
\end{figure}

\begin{figure}
\centering \epsfig{file=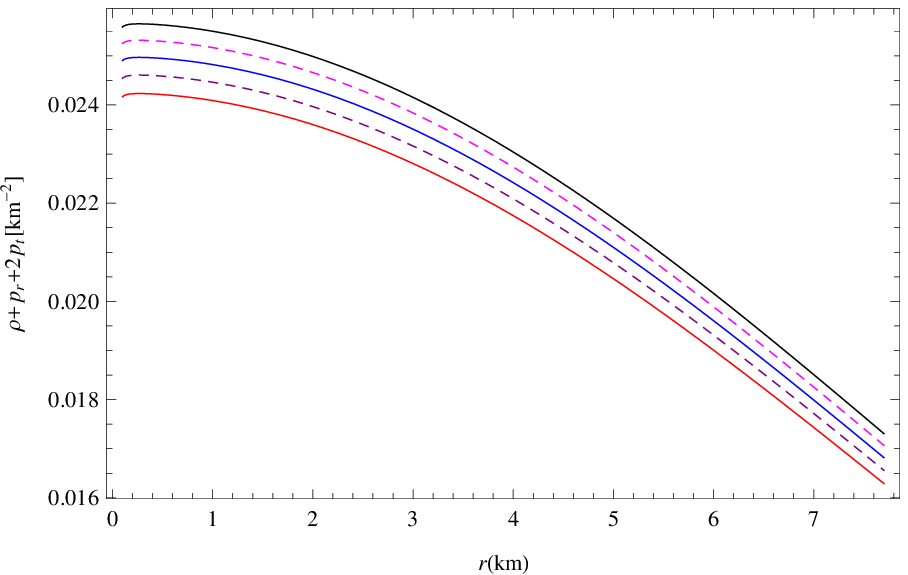, width=.32\linewidth,
height=1.50in} \epsfig{file=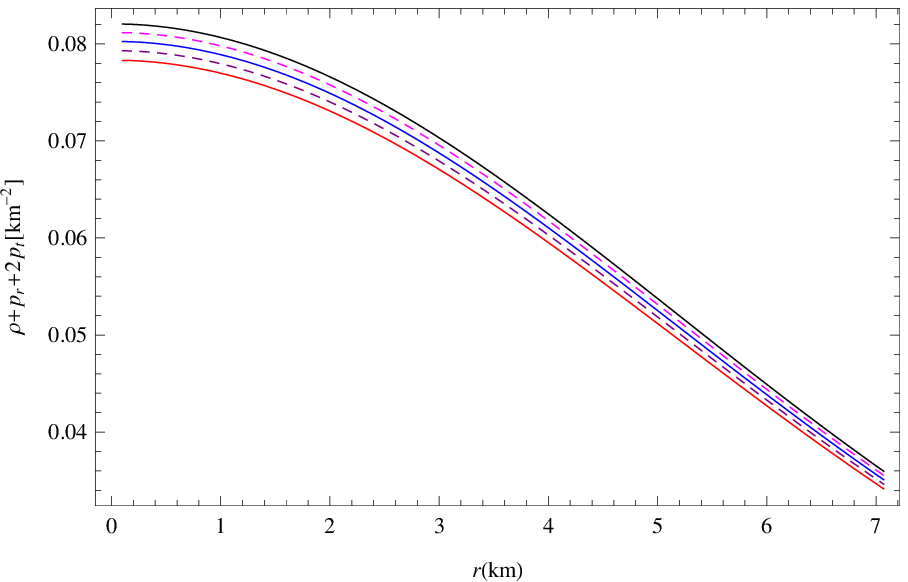, width=.32\linewidth,
height=1.50in}\epsfig{file=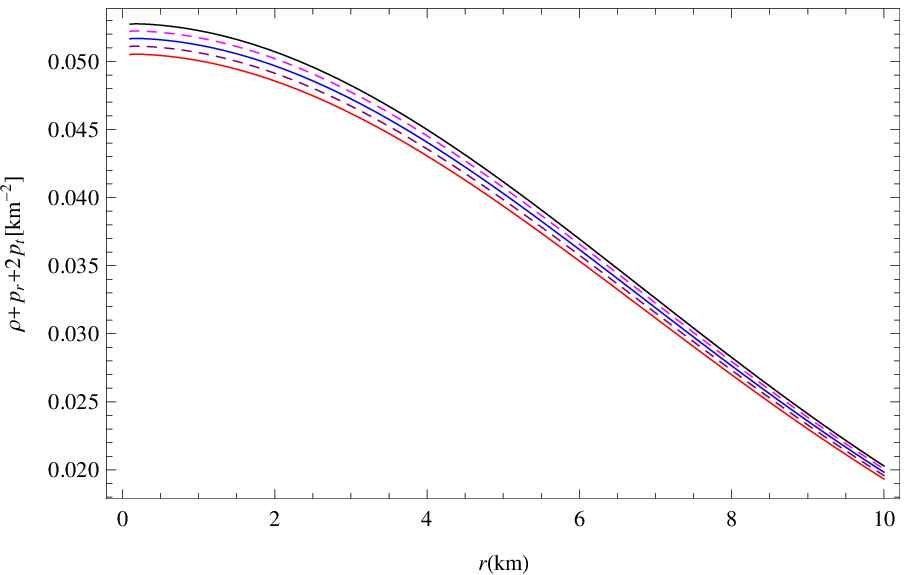, width=.32\linewidth,
height=1.50in}\caption{\label{Fig.13} Shows the energy condition for $Her X-1(Left)$, $SAX J 1808.4-3658(Middle)$, and $4U 1820-30(Right)$ with $\omega=0.01(\textcolor{red}{\bigstar})$, $\omega=0.03(\textcolor{purple}{\bigstar})$, $\omega=0.05(\textcolor{blue}{\bigstar})$, $\omega=0.07(\textcolor{magenta}{\bigstar})$, and $\omega=0.09(\textcolor{black}{\bigstar})$.}
\end{figure}
The importance of energy bounds among existing physical characteristics in explaining the presence of anisotropic compact structures is widely recognized throughout the literature. Furthermore, they serve an important role in determining the distribution of the matter content of normal as well as exotic sources inside stellar structures. Because of these energy limitations, several important conclusions can be reached. Physical attributes for the energy conditions seem to be fairly advantageous in realistically examining the matter distribution. The energy constraints have remained fundamentally vital in addressing cosmological and astrophysical challenges. Below are the expressions for $NEC$, null energy bounds, $SEC$, the strong energy bounds, $DEC$, the dominant energy bounds, and $WEC$, the week energy bounds.
\begin{eqnarray}
\begin{aligned}
NEC:\rho+p_r\geq0,~0\leq\rho+p_t;~~
WEC:\rho\geq0,~\rho+p_r\geq,~\rho+p_t\geq0,\\
SEC:0\leq\rho+p_r,~\rho+p_t\geq0,~0\leq2p_t+p_r+\rho+;~~
DEC:|p_r|<\rho,~|p_t|<\rho.\\
\end{aligned}
\end{eqnarray}
In Figs. (\ref{Fig.9}-\ref{Fig.13}), the graphical development of the energy restrictions is illustrated. The positive profiles of the energy bounds for the stars, $Her X-1$, $SAX J 1808.4-3658$, and $4U 1820-30$ for a variety of choices of $EoS$ parameter show that our solutions are realistically acceptable under RI gravity.

\subsection{Gradients and Anisotropy Analysis}

\begin{figure}
\centering \epsfig{file=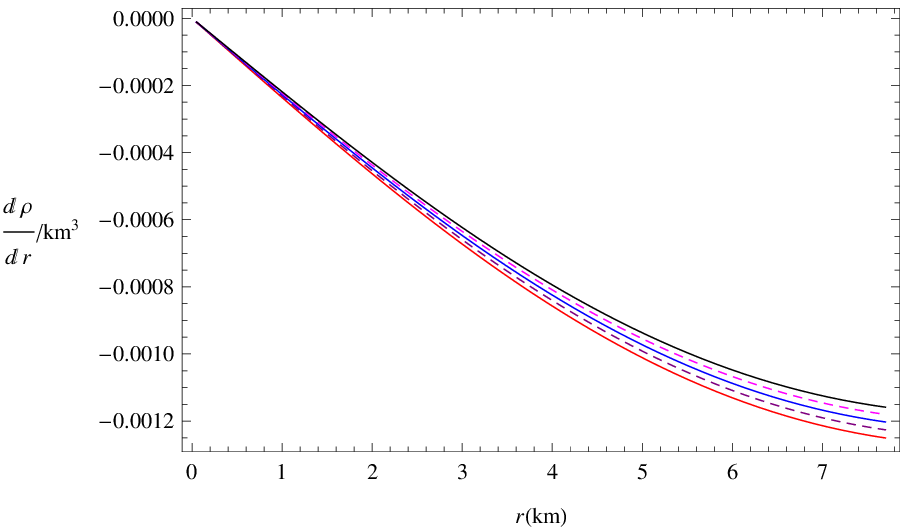, width=.32\linewidth,
height=1.50in} \epsfig{file=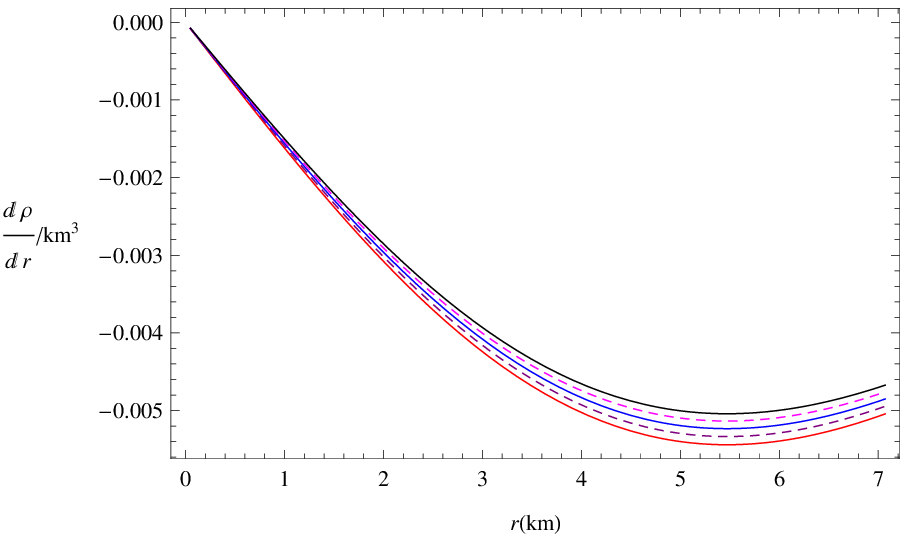, width=.32\linewidth,
height=1.50in}\epsfig{file=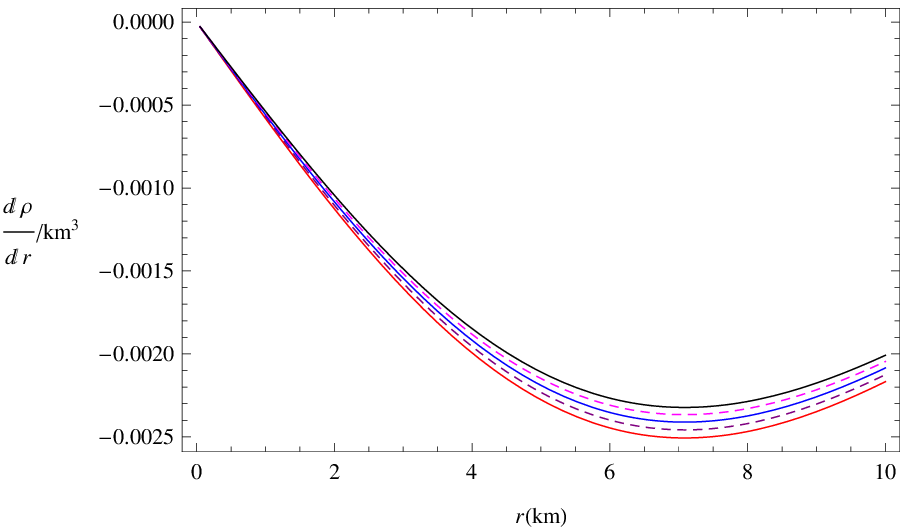, width=.32\linewidth,
height=1.50in}\caption{\label{Fig.6} Shows the derivative of energy density with respect to $r$ for $Her X-1(Left)$, $SAX J 1808.4-3658(Middle)$, and $4U 1820-30(Right)$ with $\omega=0.01(\textcolor{red}{\bigstar})$, $\omega=0.03(\textcolor{purple}{\bigstar})$, $\omega=0.05(\textcolor{blue}{\bigstar})$, $\omega=0.07(\textcolor{magenta}{\bigstar})$, and $\omega=0.09(\textcolor{black}{\bigstar})$.}
\end{figure}

\begin{figure}
\centering \epsfig{file=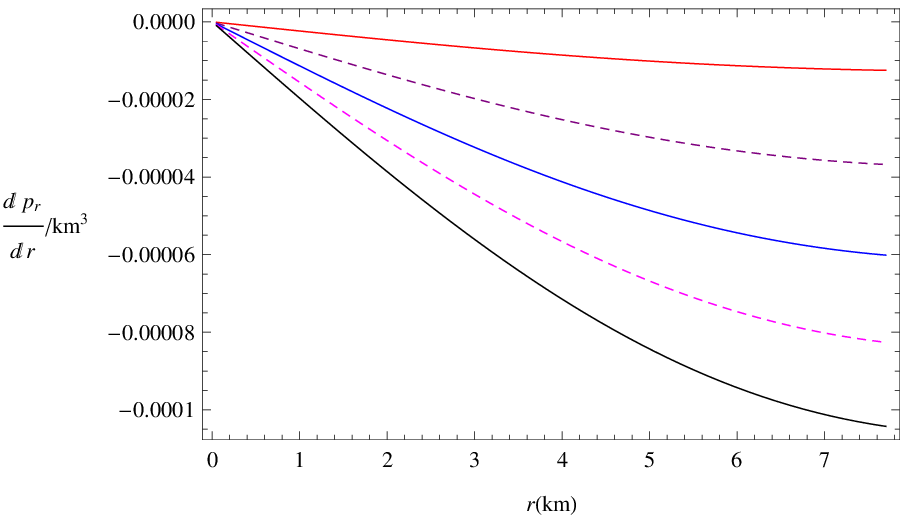, width=.32\linewidth,
height=1.50in} \epsfig{file=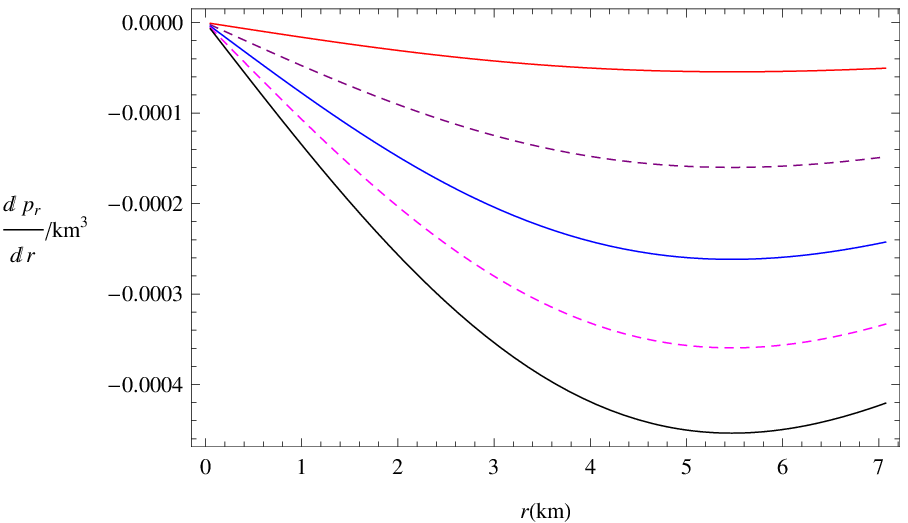, width=.32\linewidth,
height=1.50in}\epsfig{file=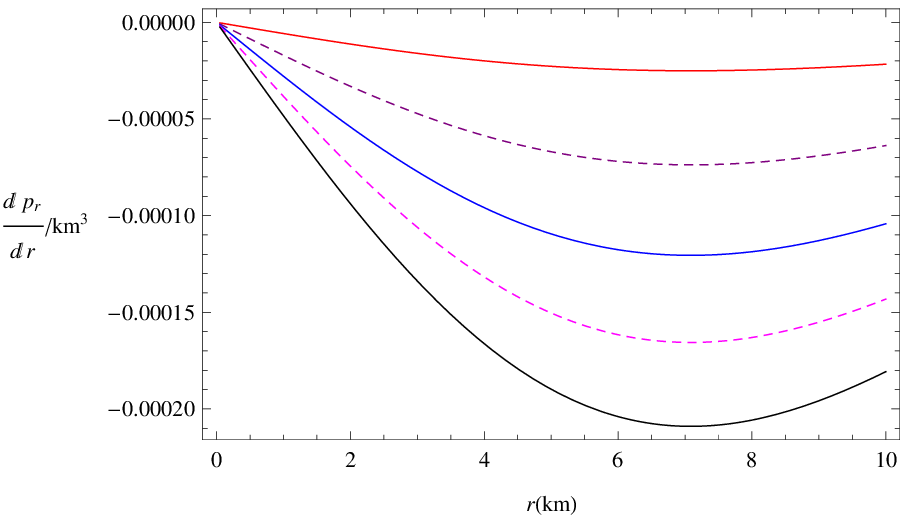, width=.32\linewidth,
height=1.50in}\caption{\label{Fig.7} Shows the derivative of radial pressure with respect to $r$ for $Her X-1(Left)$, $SAX J 1808.4-3658(Middle)$, and $4U 1820-30(Right)$ with $\omega=0.01(\textcolor{red}{\bigstar})$, $\omega=0.03(\textcolor{purple}{\bigstar})$, $\omega=0.05(\textcolor{blue}{\bigstar})$, $\omega=0.07(\textcolor{magenta}{\bigstar})$, and $\omega=0.09(\textcolor{black}{\bigstar})$.}
\end{figure}

\begin{figure}
\centering \epsfig{file=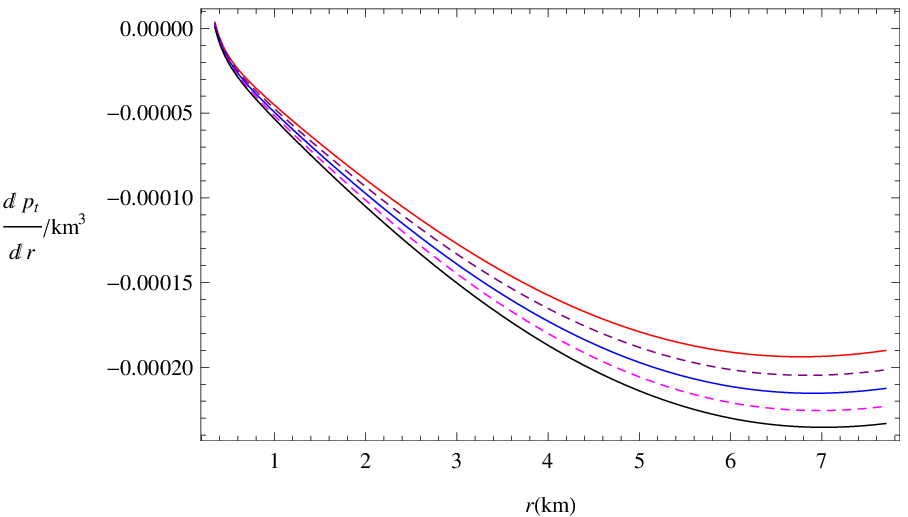, width=.32\linewidth,
height=1.50in} \epsfig{file=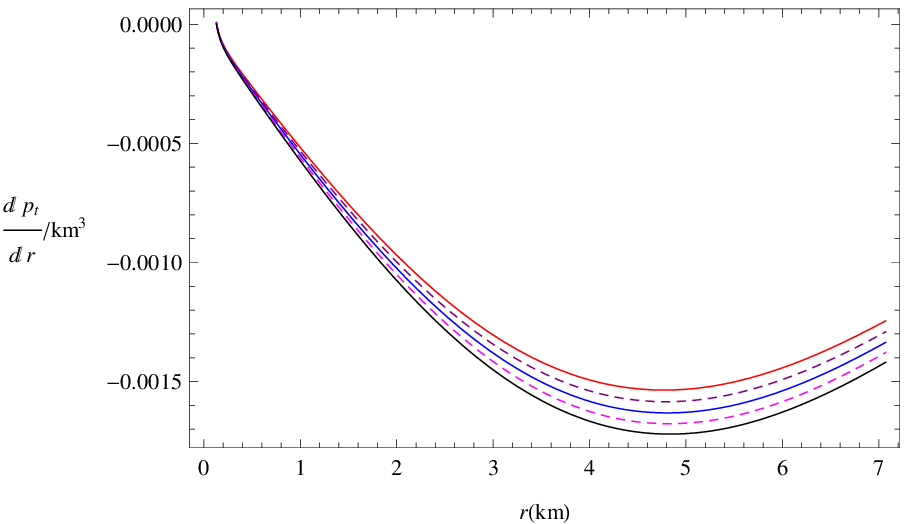, width=.32\linewidth,
height=1.50in}\epsfig{file=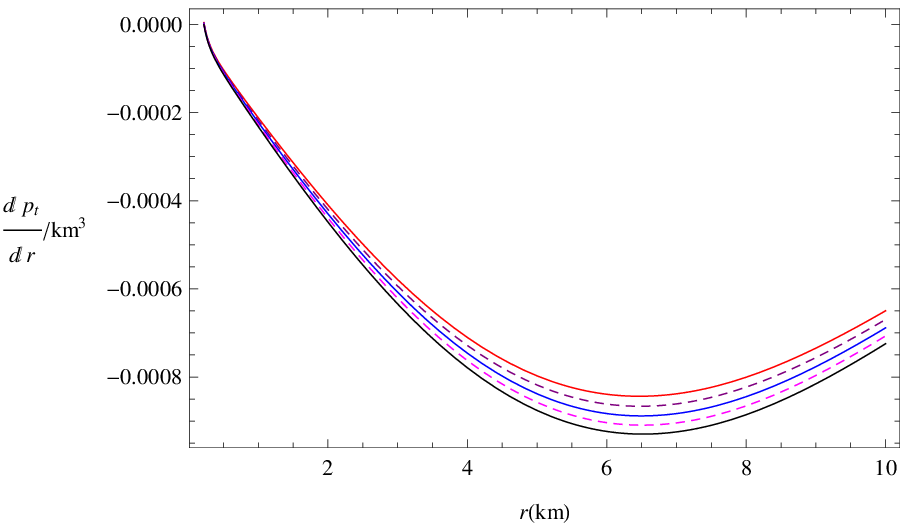, width=.32\linewidth,
height=1.50in}\caption{\label{Fig.8} Shows the derivative of tangential pressure with respect to $r$ for $Her X-1(Left)$, $SAX J 1808.4-3658(Middle)$, and $4U 1820-30(Right)$ with $\omega=0.01(\textcolor{red}{\bigstar})$, $\omega=0.03(\textcolor{purple}{\bigstar})$, $\omega=0.05(\textcolor{blue}{\bigstar})$, $\omega=0.07(\textcolor{magenta}{\bigstar})$, and $\omega=0.09(\textcolor{black}{\bigstar})$.}
\end{figure}

The total derivatives of the physical quantities $\rho$, $p_r$, and $p_t$ for compact star are described by the expressions $\frac{d\rho}{dr}$, $\frac{dp_{r}}{dr}$, and $\frac{dp_{t}}{dr}$, respectively. The graphical representations (\ref{Fig.6}-\ref{Fig.8}) of all these radial derivatives for the compact stars under discussion suggest a negative evolution, i.e., $\frac{d\rho}{dr}<0$, $\frac{dp_{r}}{dr}<0$, and $\frac{dp_{t}}{dr}<0$. It is worth noticing here that the gradients $\frac{d\rho}{dr}$, and $\frac{dp_{r}}{dr}$ $\rightarrow{0}$ at the core of the star $r=0$. This verifies the radial pressure $p_r$ attains the maximum bound as well as the density $\rho$ at the core of the star. As a result, $\rho$ and $p_r$ achieve the maximum value at $r=0$.

\begin{figure}
\centering \epsfig{file=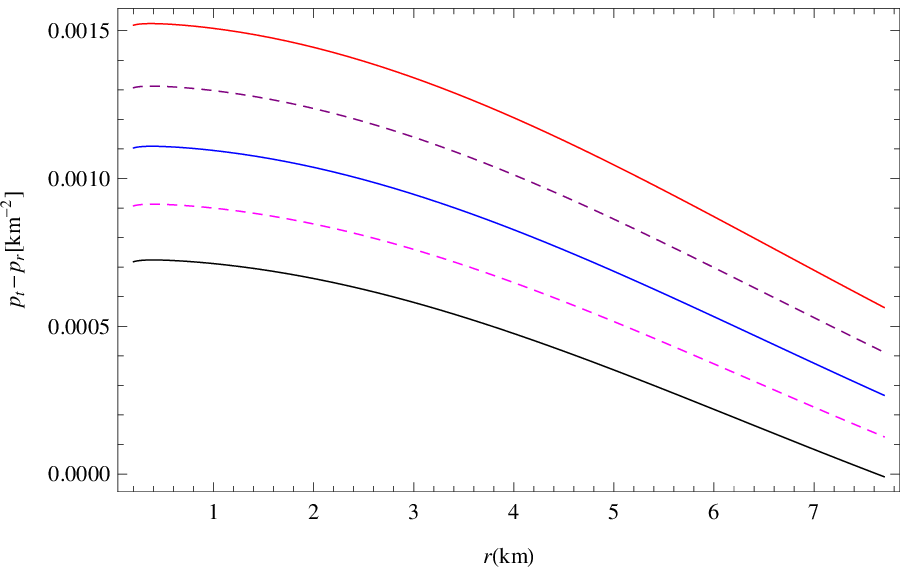, width=.32\linewidth,
height=1.50in} \epsfig{file=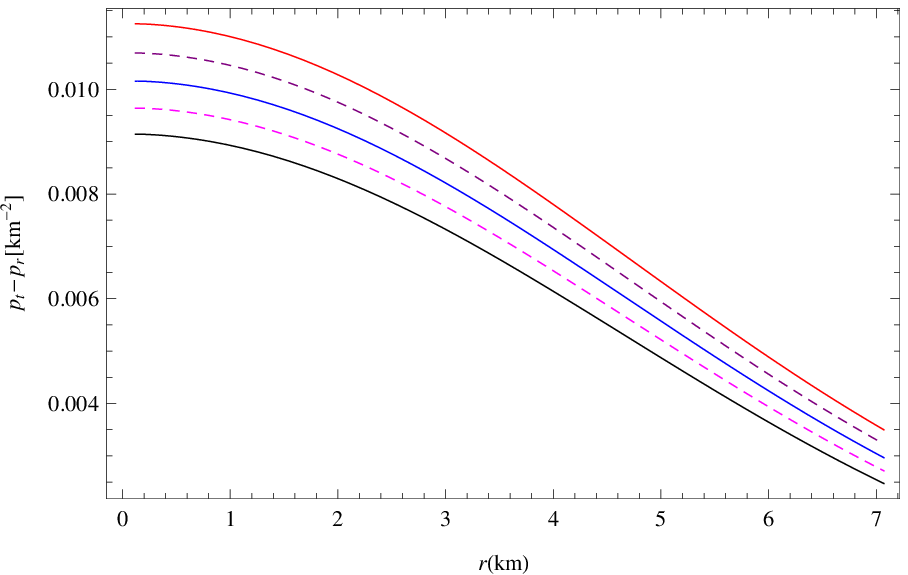, width=.32\linewidth,
height=1.50in}\epsfig{file=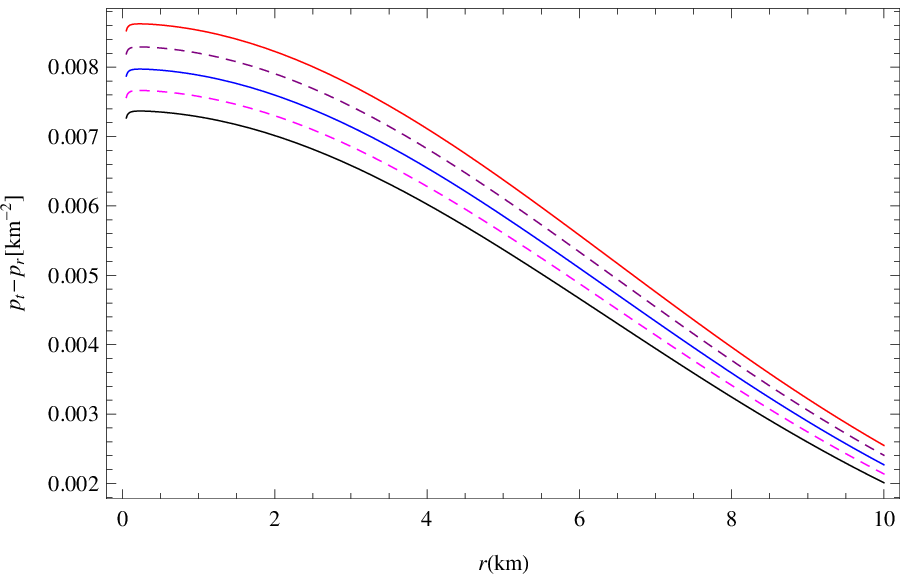, width=.32\linewidth,
height=1.50in}\caption{\label{Fig.5} Shows the anisotropy for $Her X-1(Left)$, $SAX J 1808.4-3658(Middle)$, and $4U 1820-30(Right)$ with $\omega=0.01(\textcolor{red}{\bigstar})$, $\omega=0.03(\textcolor{purple}{\bigstar})$, $\omega=0.05(\textcolor{blue}{\bigstar})$, $\omega=0.07(\textcolor{magenta}{\bigstar})$, and $\omega=0.09(\textcolor{black}{\bigstar})$.}
\end{figure}

The role of anisotropy is critical for the compact sphere modelling in order to describe the inner geometry of the relativistic stellar configuration under given conditions, and it is represented by $\Delta$, and is defined as $\Delta=p_t-p_r$. It shows the deviation between the tangential and radial stress quantities for the star configuration's anisotropic nature. If $p_t>p_r$, the anisotropy is deemed non-negative, and it is drawn outwards and seen as $\Delta>0$. If $p_r>p_t$, then, the anisotropy remains negative, resulting in $\Delta>0$, indicating that anisotropy is pulled inwards. It is evident from Fig. (\ref{Fig.5}) that anisotropy stays positive for our current investigation, i.e. $\Delta>0$, and, therefore, directed outwards.

\subsection{Stability analysis and Abreau Condition}

\begin{figure}
\centering \epsfig{file=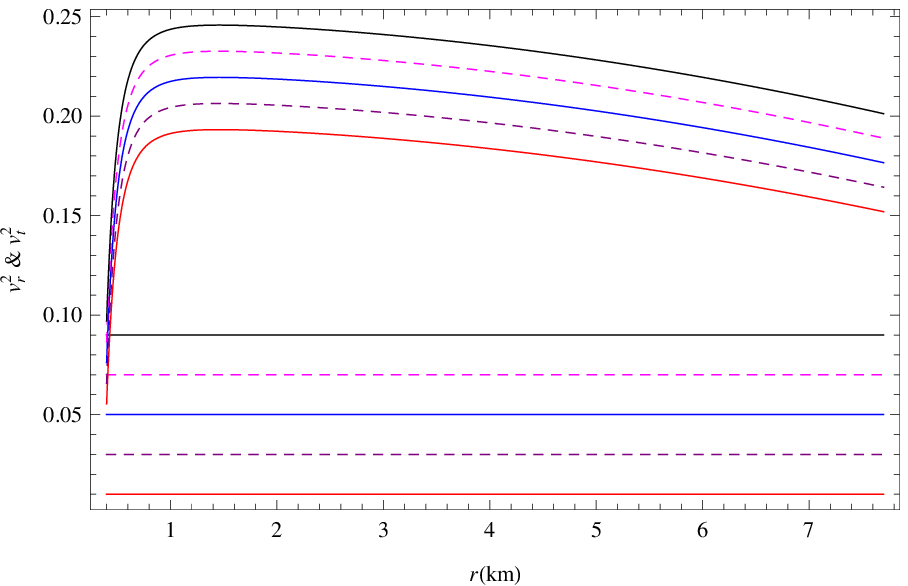, width=.32\linewidth,
height=1.50in} \epsfig{file=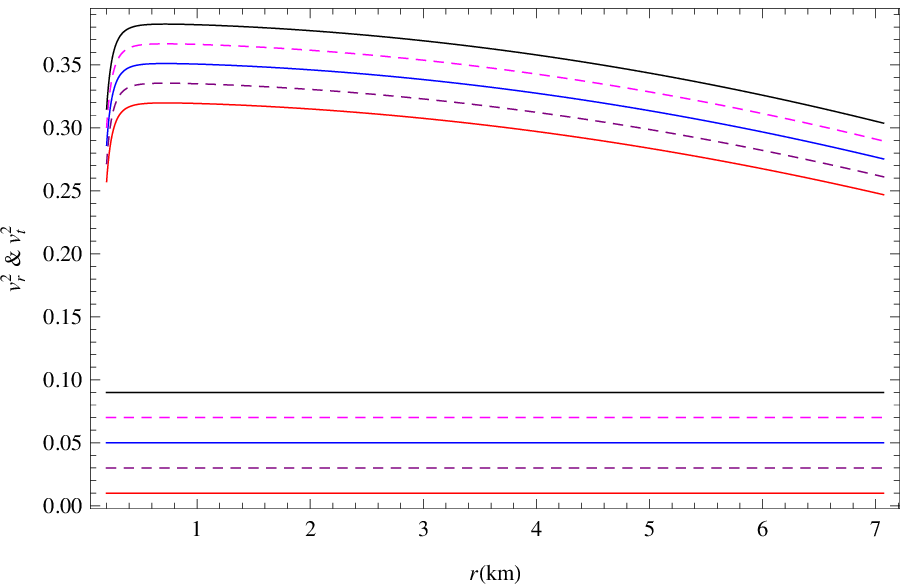, width=.32\linewidth,
height=1.50in}\epsfig{file=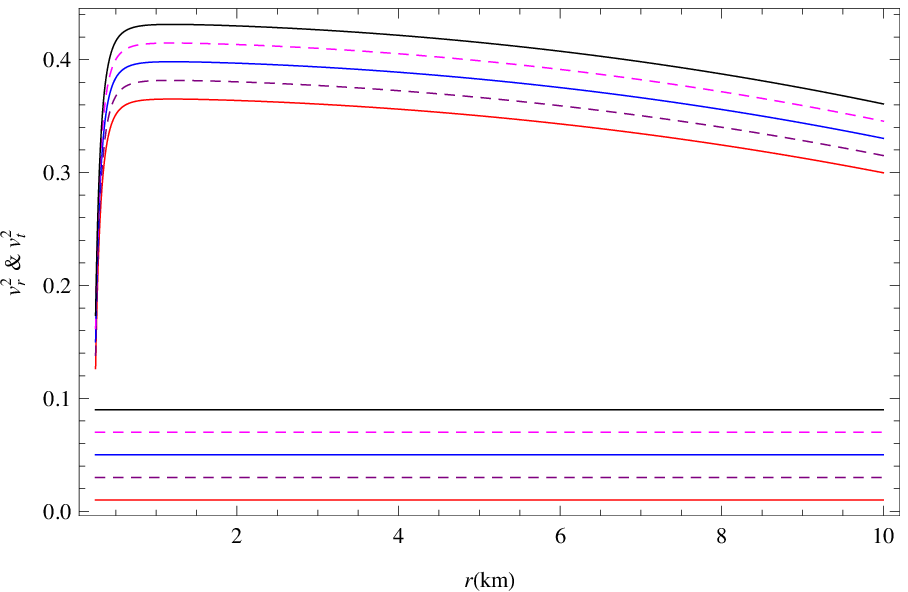, width=.32\linewidth,
height=1.50in}\caption{\label{Fig.15} Shows the speeds of sound parameters for $Her X-1(Left)$, $SAX J 1808.4-3658(Middle)$, and $4U 1820-30(Right)$ with $\omega=0.01(\textcolor{red}{\bigstar})$, $\omega=0.03(\textcolor{purple}{\bigstar})$, $\omega=0.05(\textcolor{blue}{\bigstar})$, $\omega=0.07(\textcolor{magenta}{\bigstar})$, and $\omega=0.09(\textcolor{black}{\bigstar})$.}
\end{figure}

\begin{figure}
\centering \epsfig{file=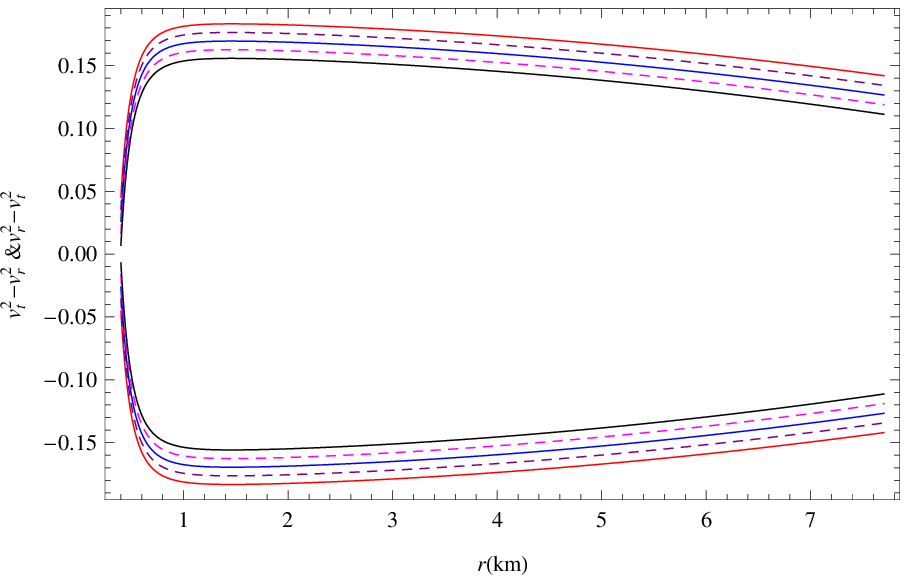, width=.32\linewidth,
height=1.50in} \epsfig{file=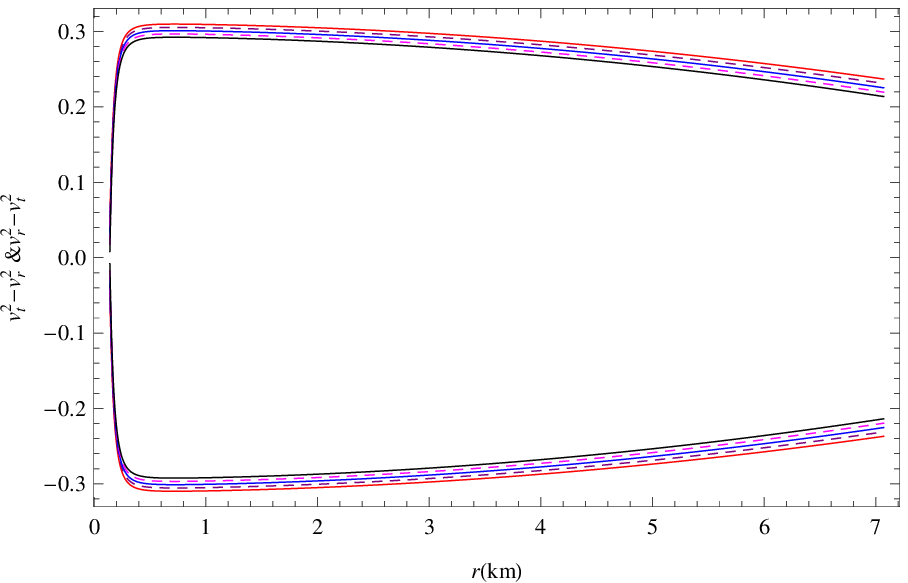, width=.32\linewidth,
height=1.50in}\epsfig{file=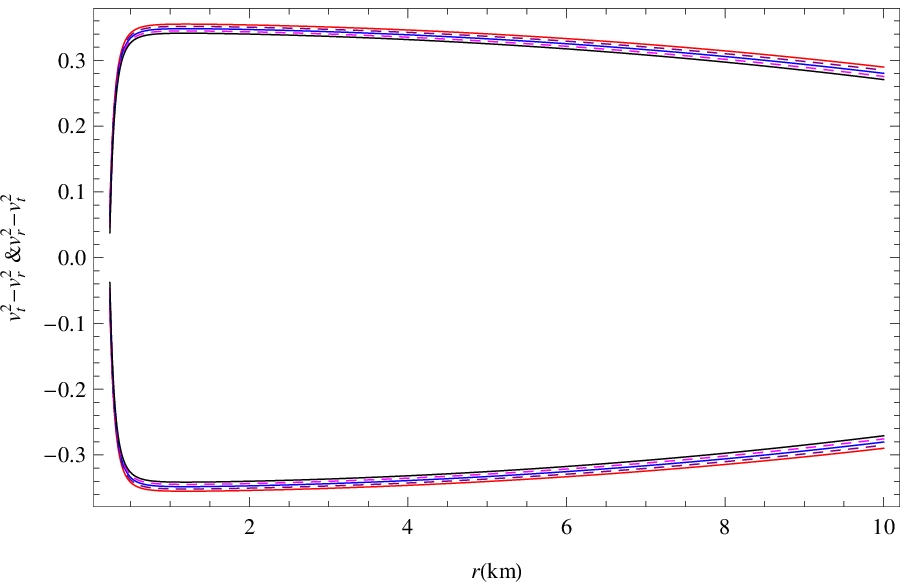, width=.32\linewidth,
height=1.50in}\caption{\label{Fig.16} Shows the Abreu condition for $Her X-1(Left)$, $SAX J 1808.4-3658(Middle)$, and $4U 1820-30(Right)$ with $\omega=0.01(\textcolor{red}{\bigstar})$, $\omega=0.03(\textcolor{purple}{\bigstar})$, $\omega=0.05(\textcolor{blue}{\bigstar})$, $\omega=0.07(\textcolor{magenta}{\bigstar})$, and $\omega=0.09(\textcolor{black}{\bigstar})$.}
\end{figure}

The sound speeds related with the radial and transversal components with their corresponding representations of $v^{2}_{r}$ A and $v^{2}_{t}$  establish  the stability of the compact star solutions obtained in our investigations. For this purpose, The conditions $0\leq{v^{2}_{t}}=\frac{dp_{t}}{d\rho}\leq1$ and $0\leq{v^{2}_{r}}=\frac{dp_{r}}{d\rho}\leq1$ must be achieved
\cite{Herrera}. The related plots of sound speeds in Figs. (\ref{Fig.15}) for different values of $EoS$ parameter demonstrate that the radial and transversal sound speeds for the stellar candidates $Her X-1$, $SAX J 1808.4-3658$, and $4U 1820-30$ remain within the necessary stability restrictions.
The limitations of both the radial and transversal sound velocities are validated for all compact star candidates. The approximation of the steady and fragile epochs inside the anisotropic matter configurations can be discussed by modifying the sound speeds propagations, which have the expression $v^{2}_{t}-v^{2}_{r}$ endorsing the constraint $0<|v^{2}_{t}-v^{2}_{r}|<1$, and this condition is referred to as the Abreu condition depicted in Fig.(\ref{Fig.16}). It shows that the for the compact stars discussed under RI gravity, total stability may be achieved.

\begin{figure}
\centering \epsfig{file=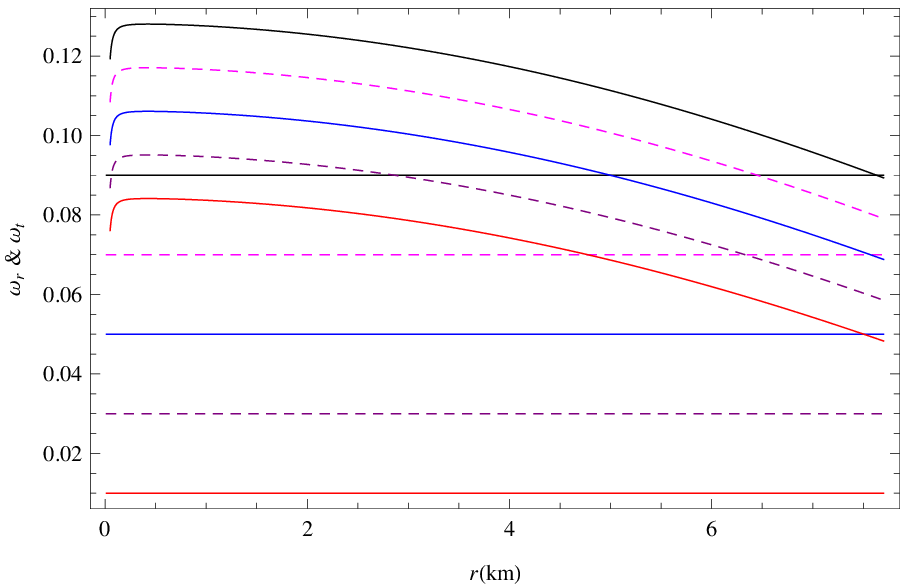, width=.32\linewidth,
height=1.50in} \epsfig{file=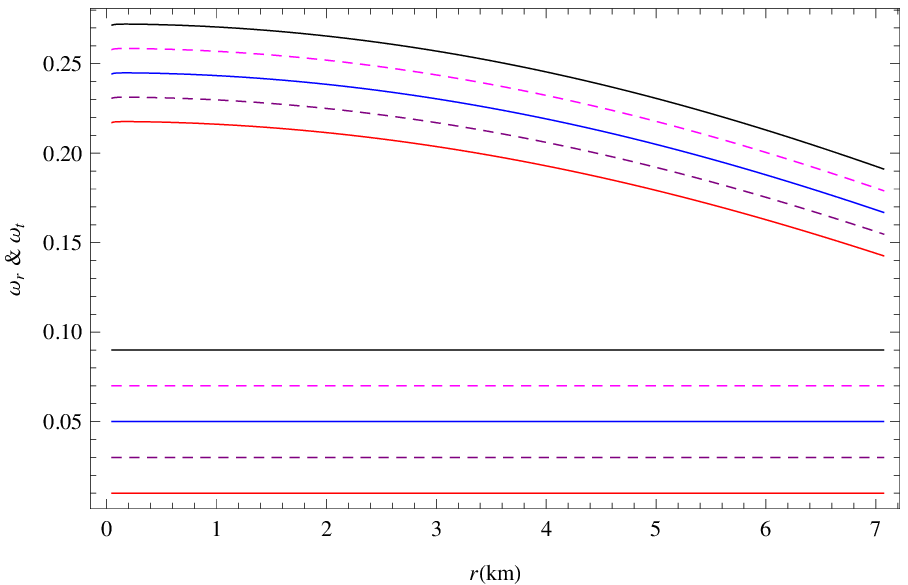, width=.32\linewidth,
height=1.50in}\epsfig{file=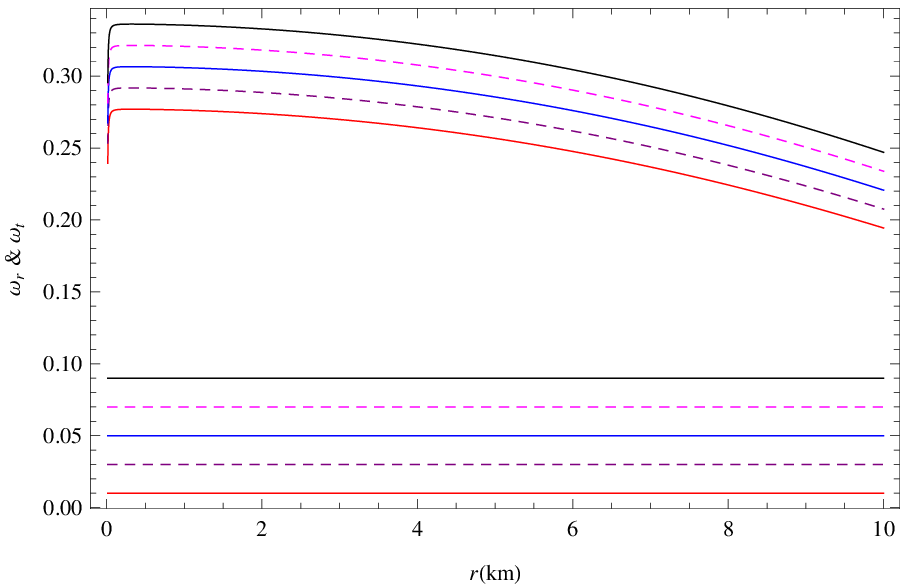, width=.32\linewidth,
height=1.50in}\caption{\label{Fig.14} Shows the equations of state parameters for $Her X-1(Left)$, $SAX J 1808.4-3658(Middle)$, and $4U 1820-30(Right)$ with $\omega=0.01(\textcolor{red}{\bigstar})$, $\omega=0.03(\textcolor{purple}{\bigstar})$, $\omega=0.05(\textcolor{blue}{\bigstar})$, $\omega=0.07(\textcolor{magenta}{\bigstar})$, and $\omega=0.09(\textcolor{black}{\bigstar})$.}
\end{figure}

\subsection{Equation of State Parameter}

The development of compact star emergence can be predicted using $EoS$ parameter. Furthermore, $EoS$ is a crucial component of any astrophysical model. The pressure terms $p_r$ and $p_t$ are used to calculate the $EoS$ for the anisotropic matter distribution. The components of $EoS$ for studying stellar configuration are $\omega_r$ and $\omega_t$, which are mathematically written as
\begin{equation}
 \omega_r=\frac{p_r}{\rho_e}, ~~~~~~  \omega_t=\frac{p_t}{\rho_e}.
\end{equation}
The components of the $EoS$ exhibit monotonically decreasing behavior with increasing radii and is always below 1, which can be witnessed in Fig. \ref{Fig.14}. Furthermore,the positive character of both $EoS$ components, $\omega_r$ and $\omega_t$, is seen which fulfills the requirement $0 \leq\omega_r$  and $\omega_t<1$ for the viable stellar solutions.

\subsection{Mass-Radius relation, Compactness, and Redshift Analysis}

\begin{figure}
\centering \epsfig{file=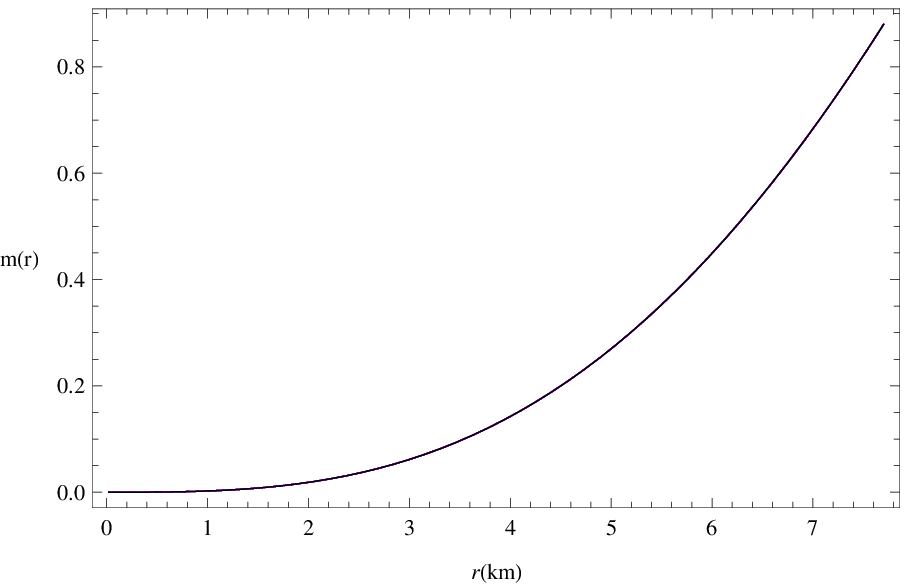, width=.32\linewidth,
height=1.50in} \epsfig{file=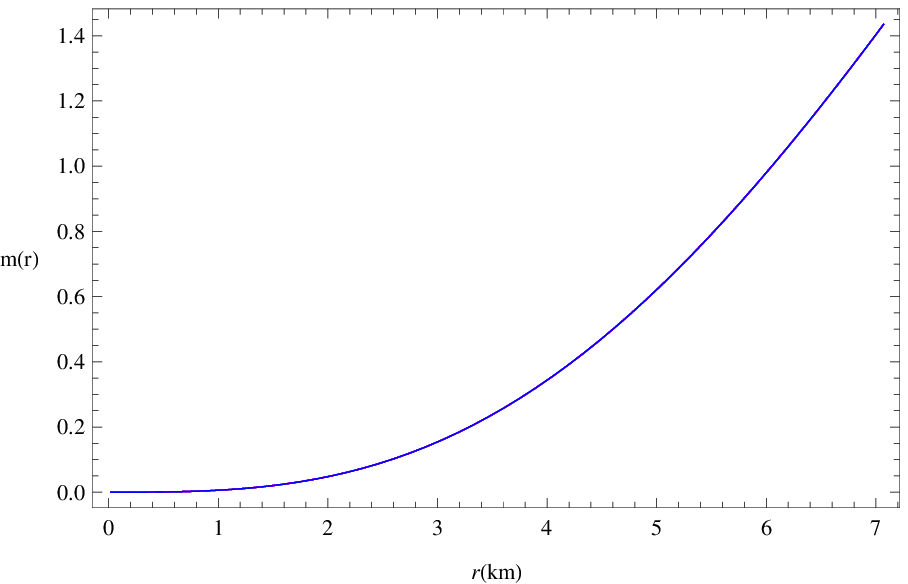, width=.32\linewidth,
height=1.50in}\epsfig{file=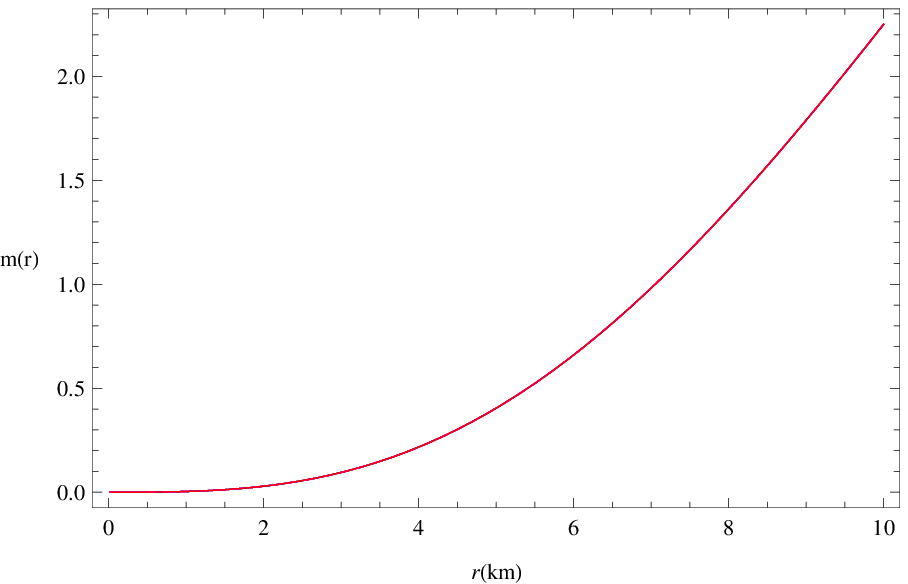, width=.32\linewidth,
height=1.50in}\caption{\label{Fig.17} Shows the mass function for $Her X-1(Left)$, $SAX J 1808.4-3658(Middle)$, and $4U 1820-30(Right)$ with $\omega=0.01(\textcolor{red}{\bigstar})$, $\omega=0.03(\textcolor{purple}{\bigstar})$, $\omega=0.05(\textcolor{blue}{\bigstar})$, $\omega=0.07(\textcolor{magenta}{\bigstar})$, and $\omega=0.09(\textcolor{black}{\bigstar})$.}
\end{figure}

\begin{figure}
\centering \epsfig{file=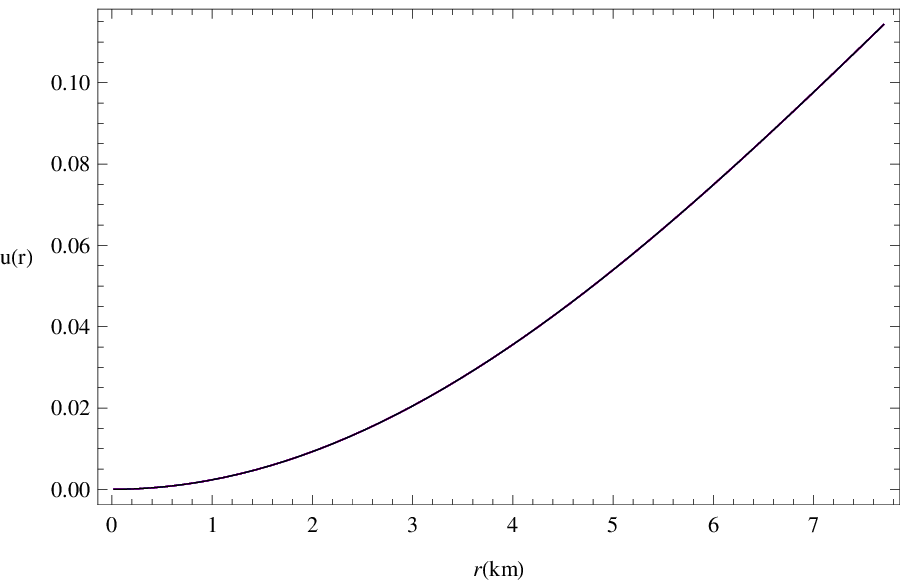, width=.32\linewidth,
height=1.50in} \epsfig{file=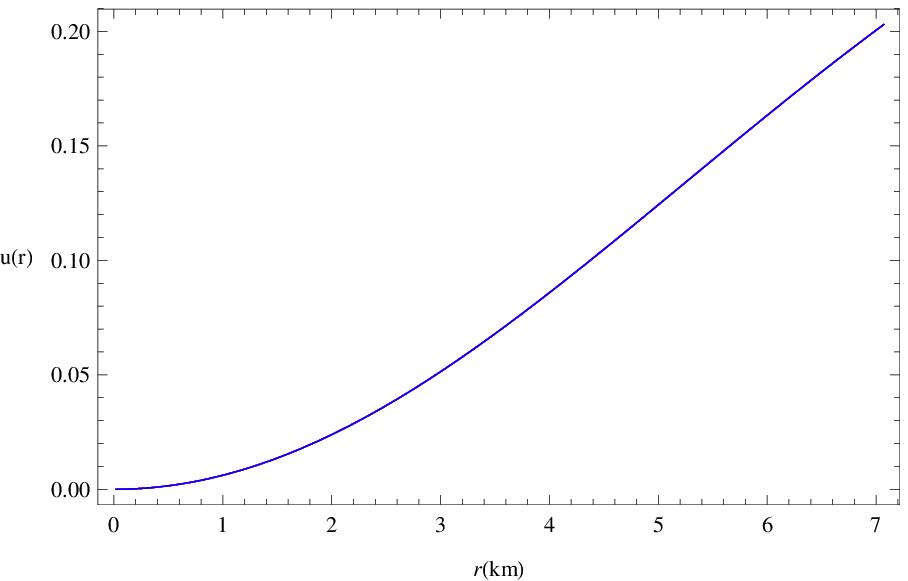, width=.32\linewidth,
height=1.50in}\epsfig{file=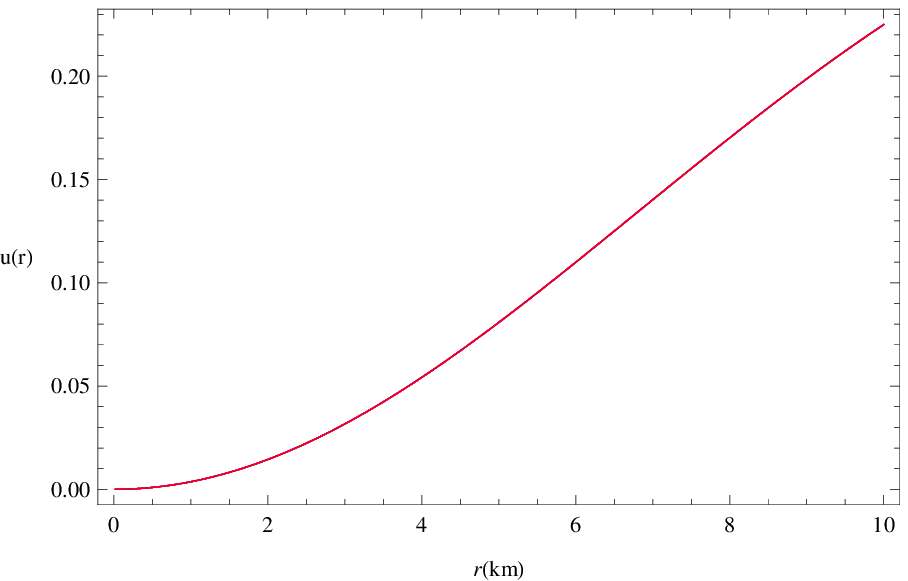, width=.32\linewidth,
height=1.50in}\caption{\label{Fig.18} Shows the compactness function for $Her X-1(Left)$, $SAX J 1808.4-3658(Middle)$, and $4U 1820-30(Right)$ with $\omega=0.01(\textcolor{red}{\bigstar})$, $\omega=0.03(\textcolor{purple}{\bigstar})$, $\omega=0.05(\textcolor{blue}{\bigstar})$, $\omega=0.07(\textcolor{magenta}{\bigstar})$, and $\omega=0.09(\textcolor{black}{\bigstar})$.}
\end{figure}

\begin{figure}
\centering \epsfig{file=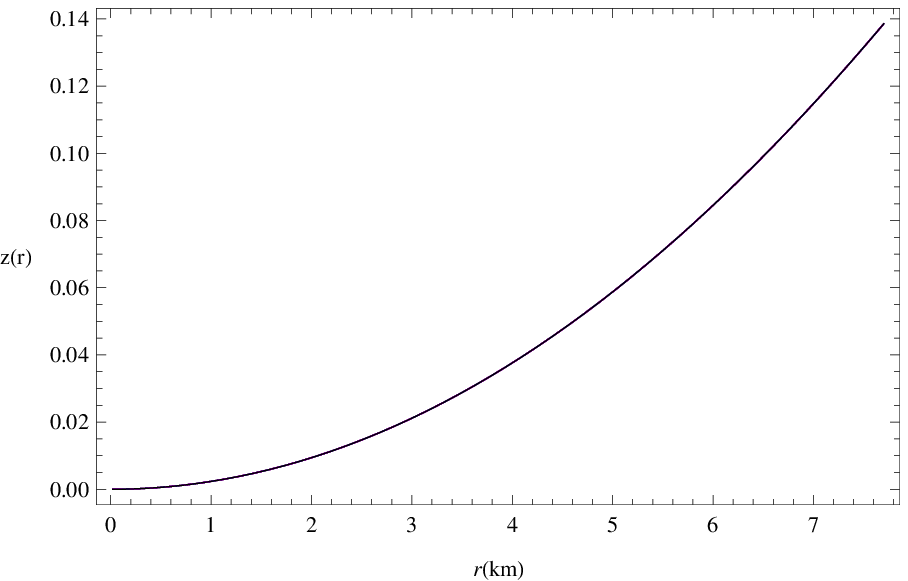, width=.32\linewidth,
height=1.50in} \epsfig{file=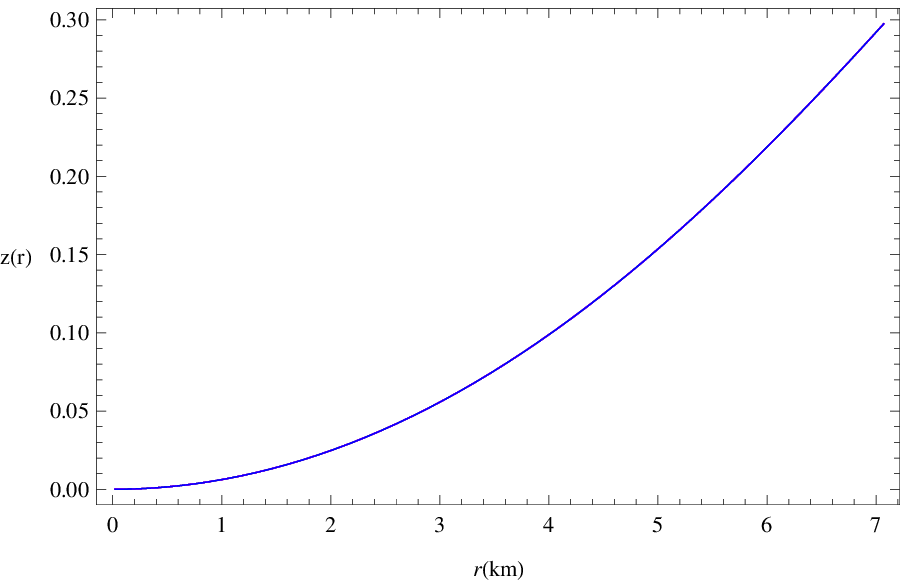, width=.32\linewidth,
height=1.50in}\epsfig{file=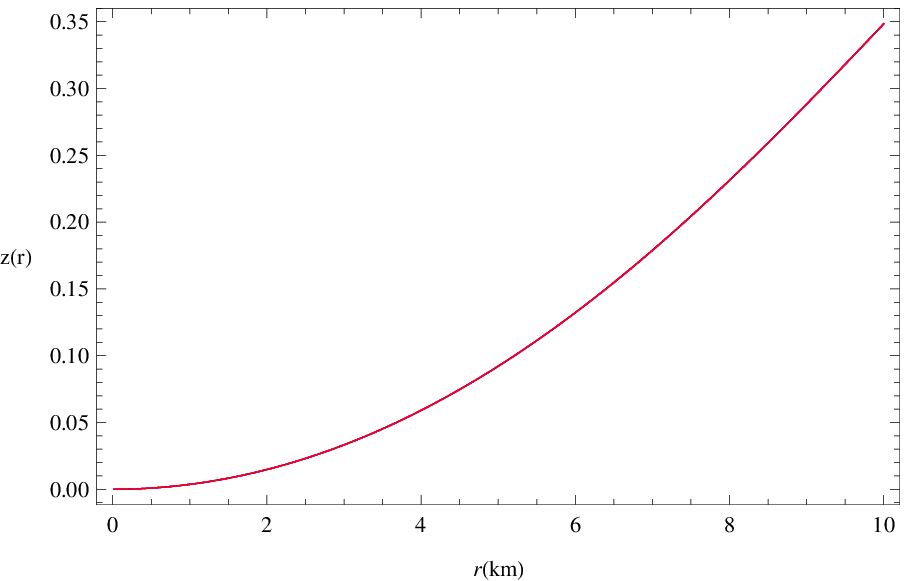, width=.32\linewidth,
height=1.50in}\caption{\label{Fig.19} Shows the redshift function for $Her X-1(Left)$, $SAX J 1808.4-3658(Middle)$, and $4U 1820-30(Right)$ with $\omega=0.01(\textcolor{red}{\bigstar})$, $\omega=0.03(\textcolor{purple}{\bigstar})$, $\omega=0.05(\textcolor{blue}{\bigstar})$, $\omega=0.07(\textcolor{magenta}{\bigstar})$, and $\omega=0.09(\textcolor{black}{\bigstar})$.}
\end{figure}

Radii $r$ dependent mass function of a compact star, $m(r)$ is determined by employing the integral as follows
\begin{equation}\label{mass}
m(r)=4\int_{0}^{r}\pi\acute{r^2}\rho{d}\acute{r}.
\end{equation}
The mass-radius graph in Fig. (\ref{Fig.17}) shows that the mass is exactly correlated to the radius $r$, so that as $r\rightarrow0$, $m(r)\rightarrow0$,  indicating that the correct behavior of mass at the star's core. Also, as calculated by Bhomer and Harko \cite{boht}, the mass-radius ratio must remain $\frac{2M}{r}\leq\frac{8}{9}$ , which in also validated in our study. The stellar structure's compactness $\mathcal{K}(r)$ is given by the integral
\begin{equation}\label{Compactness}
\mathcal{K}(r)=\frac{4\pi}{r}\int_{0}^{r}\rho\acute{r^2}{d}\acute{r},
\end{equation}
and plotted in Fig. (\ref{Fig.18}).
The redshift $\Re_{S}$ function is given as
\begin{equation}\label{RedShift}
\Re_{S}+1=[1-2\mathcal{K}(r)]^{\frac{-1}{2}}.
\end{equation}
Fig. (\ref{Fig.19}) shows its graphical depiction and it is concluded that the numerical approximation of $\Re_{S}$ is still within the required threshold  $\Re_{S}\leq2$, indicating that our worked out solutions in RI gravity are indeed viable.

\section{Results and discussions}

This study is focussed to explore anisotropic compact structures, using a new alternative theory termed as RI Gravity \cite{Amendola}. The related field equations of the RI gravity for compact stars are solved using certain assumptions. We establish our calculations based on the statistics related to the $Her X-1$, $SAX J 1808.4-3658$, and $4U 1820-30$ as strange star candidates, by choosing appropriate $EoS$ parametric values. Our current investigation is focused on the feasibility of the existence of quintessence anisotropic compact stars. The gravitational breakdown for the emerging universe at diverse epochs has now been examined by integrating spacetime symmetries with the exclusively  matched  Schwarzschild's geometry. Following are some of the primary findings from current analysis along with some key points.
\begin{itemize}
\item The terms $\rho$, $p_t$, $p_r$ are critical physical factors for the creation of stellar formations. The related plots and the statistics connected to the strange star candidates under our investigation indicate that the energy density near the star core reaches the maximum value, indicating the star's high-density nature. Furthermore, the tangential and radial pressure terms remain positive and show monotonically decreasing behavior towards the boundary of the star. These patterns also provide evidence for the presence of anisotropic matter distribution free from singularities for the RI gravity under consideration. Additionally, the plots of the quintessence energy density show a negative trend, indicating that our stellar RI gravity model supports quintessence.
\item The importance of energy restrictions in the literature on compact stellar structures is fairly clear. All the related plots show that energy constraints are satisfies proving that our obtained solutions are physically viable under RI gravity. The gradient profiles $\frac{d\rho}{dr}$, $\frac{d p_{r}}{dr}$ and $\frac{d p_{t}}{dr}$  of the compact stars are shown in the respective plots. It is indicated that the first ordere derivative is evolving negatively which verifies the upper bound of the radial pressure $p_r$ with respect to the density $\rho$. As a result, at $r=0$, $\rho$ and $p_r$ attain the maximum value.
\item The associated sound speed graphs are illustrated by the relevant figures, which demonstrate that the radial, as well as the  transversal sound speeds for all presented stars, are confined within the required limits of consistency. The graphical developments demonstrate that the condition $0<|v^{2}_{st}-v^{2}_{sr}|<1$ is met for the stable behavior of a compact star.
\item  The $EoS$ parameters for the stars $Her X-1$, $SAX J 1808.4-3658$, and $4U 1820-30$ have also been plotted for the various values. It is concluded that the constraints $\omega_r$ and $\omega_t$ of $EoS$ remain positive in the stellar interior  and are within the stability bounds  $0\leq\omega_r$,A and $\omega_t <1$ under RI gravity.
\item The anisotropy $\Delta$ with respect to $r$ of the stars under observation shows a positive increasing behavior as illustrated in their associated graphs, indicating repelling anisotropic forces accommodated by a dense matter source.

\item Finally, the compactness parameter, mass function, and redshift are also plotted. It can be observed that $\mu(r)\leq 0.30$  and has an increasing tendency, whereas $m(r)$ also has an increasing character. Further, $Z_s$ has a decreasing attribute and is always $Z_s\leq 5$, which is associated with the  stability of the compact structures .
\end{itemize}

\section*{Appendix (\textbf{I})}
\begin{eqnarray*}
\chi _1&&=-\frac{2 r^2}{r^2 (A-B)+e^{A r^2}-1}+\frac{3}{B \left(r^2 (B-A)+3\right)}+\frac{1}{B r^2 (B-A)-2 A+B},\\
\chi _2&&=-72 A^3 r^4+143 A^2 B r^4+114 A^2 r^2-88 A B^2 r^4-117 A B r^2+72 A+30 B^2 r^2-45 B,\\
\chi _3&&=10 A^4 r^6-29 A^3 B r^6+30 A^2 B^2 r^6-13 A B^3 r^6+2 B^4 r^6+17 B^3 r^4,\\
\chi _4&&=-2 B r^2 e^{2 A r^2} \left(r^2 (B-A)+3\right) \left(r^2 \left(2 A^2-3 A B+B^2\right)-8 A+5 B\right),\\
\chi _5&&=\frac{2 r^2 \left(r^2 \left(2 A^2-3 A B+B^2\right)-8 A+5 B\right)}{B \left(r^2 (B-A)+3\right)^3},\\
\chi _6&&=3 A^4 r^3 \left(B r^2+2\right)^2-A^3 r \left(B r^2+2\right) \left(B r^2 (B r (r+6)+25)-2\right)+A^2 B r (B r (r (B r (r (B (2 r+3)+4)+35)\\&&+61)+14)-8)-A B^2 (r (B r (r (B (r (B r+5)+8)+3)+45)-3)+4)+B^3 \left(B r \left(B r^2+r+11\right)+2\right),\\
\chi _7&&=\frac{\frac{1}{B r^2 (A-B)+2 A-B}-\frac{1}{B \left(r^2 (B-A)+3\right)}}{2 r^2}+\frac{1}{r^2 (A-B)+e^{A r^2}-1},\\
\chi _8&&=\frac{r^2 \left(2 A^2-3 A B+B^2\right)-8 A+5 B}{B \left(r^2 (B-A)+3\right)^3},\\
\chi _9&&=r^4 \left(3 A^2-10 A B-B^2\right)+e^{A r^2} \left(r^6 (A-B) \left(3 A^2-8 A B+B^2\right)+2 A r^8 (A-B)^2 (A+B)-2 r^4 (A-B)\right. \\&& \left.(17 A-B)+r^2 (27 A-7 B)+8\right)-8 A^2 r^8 (A-B)^2+2 A r^6 (A-B) (9 A-B)+2 r^2 (4 B-11 A)+e^{3 A r^2}\\&&\times \left(r^2 (A+B)+2\right)+e^{2 A r^2} \left(2 A r^6 (A-B) (A+B)+2 A r^4 (A-3 B)-2 r^2 (3 A+B)-7\right)-3,\\
\chi _{10}&&=\frac{\alpha  e^{2 A r^2}}{r^2 (A-B)+e^{A r^2}-1}+\frac{A-2 B}{r^2}+B (A-B).
\end{eqnarray*}
\begin{eqnarray*}
\Psi _1&&=\frac{2 r^2 \left(r^2 \left(2 A^2-3 A B+B^2\right)-8 A+5 B\right)}{B \left(r^2 (B-A)+3\right)^3}+\frac{1}{r \left(B (B r+1)-A \left(B r^2+2\right)\right)^4},\\
\Psi _2&&=3 A^4 r^3 \left(B r^2+2\right)^2-A^3 r \left(B r^2+2\right) \left(B r^2 (B r (r+6)+25)-2\right)+A^2 B r (B r (r (B r (r (B (2 r+3)+4)\\&&+35)+61)+14)-8)-A B^2 (r (B r (r (B (r (B r+5)+8)+3)+45)-3)+4)+B^3 \left(B r \left(B r^2+r+11\right)+2\right),\\
\Psi _3&&=-\frac{2 r^2}{r^2 (A-B)+e^{A r^2}-1}+\frac{3}{B \left(r^2 (B-A)+3\right)}+\frac{1}{B r^2 (B-A)-2 A+B},\\
\Psi _4&&=10 A^4 r^6-29 A^3 B r^6-72 A^3 r^4+30 A^2 B^2 r^6-2 B r^2 e^{2 A r^2} \left(r^2 (B-A)+3\right) \left(r^2 \left(2 A^2-3 A B+B^2\right)-8 A+5\right.\nonumber\\&&\times\left. B\right)+143 A^2 B r^4+114 A^2 r^2-13 A B^3 r^6-88 A B^2 r^4-117 A B r^2+72 A+2 B^4 r^6+17 B^3 r^4+30 B^2 r^2-45 B,\\
\Psi _5&&=\frac{\frac{1}{B r^2 (A-B)+2 A-B}-\frac{1}{B \left(r^2 (B-A)+3\right)}}{2 r^2}+\frac{1}{r^2 (A-B)+e^{A r^2}-1},\\
\Psi _6&&=\frac{r^2 \left(2 A^2-3 A B+B^2\right)-8 A+5 B}{B \left(r^2 (B-A)+3\right)^3}+\frac{1}{\left(r^2 (A-B)+e^{A r^2}-1\right)^4},\\
\Psi _7&&=r^4 \left(3 A^2-10 A B-B^2\right)-8 A^2 r^8 (A-B)^2+2 A r^6 (A-B) (9 A-B)+2 r^2 (4 B-11 A)+e^{3 A r^2} \\&&\times\left(r^2 (A+B)+2\right)+e^{2 A r^2}\left(2 A r^6 (A-B) (A+B)+2 A r^4 (A-3 B)-2 r^2 (3 A+B)-7\right)-3,\\
\Psi _8&&=r^6 (A-B) \left(3 A^2-8 A B+B^2\right)+2 A r^8 (A-B)^2 (A+B)-2 r^4 (A-B) (17 A-B)+r^2 (27 A-7 B)+8
\end{eqnarray*}

\end{document}